\documentclass[a4paper]{scrartcl}
\pdfoutput=1
\usepackage[utf8]{inputenc}
\usepackage[T1]{fontenc}
\usepackage{lmodern}
\usepackage{exscale} % To scale mathematical symbols correctly while using T1
\usepackage{amsmath,amsthm,amssymb}
\usepackage{dsfont}
\usepackage[british]{babel}
\usepackage{enumitem}
\usepackage{microtype} % Microtypography!
\usepackage{hyperref}
\usepackage{authblk}
\usepackage{xcolor}

\usepackage[sort&compress,numbers]{natbib}
\DeclareMathOperator{\Or}{O}
\newcommand{\D}{\mathrm{d}}
\newcommand{\I}{\mathrm{i}}
\newcommand{\E}{\mathrm{e}}

\newcommand{\vect}[1]{\boldsymbol{\mathbf{#1}}} % boldface upright vectors

\newcommand{\ul}{\underline} % shorthand for underlined indices

\numberwithin{equation}{section}

\title{Post-Newtonian Hamiltonian description of 
an atom in a weak gravitational field}

\author[1,a]{\textsc{Philip K. Schwartz}}
\author[1,2,b]{\textsc{Domenico Giulini}}
\affil[1]{Institute for Theoretical Physics,
	Leibniz University Hannover, \par
	Appelstra{\ss}e 2, 30167 Hannover, Germany}
\affil[2]{Center of Applied Space Technology and Microgravity,
	University of Bremen, \par
	Am Fallturm 1, 28359 Bremen, Germany}
\affil[a]{\normalfont\texttt{\href{mailto:philip.schwartz@itp.uni-hannover.de}{philip.schwartz@itp.uni-hannover.de}}}
\affil[b]{\normalfont\texttt{\href{mailto:giulini@itp.uni-hannover.de}{giulini@itp.uni-hannover.de}}}

\date{}

\begin{document}
\maketitle

\begin{abstract}
\noindent
We extend the systematic calculation of an approximately 
relativistic Hamiltonian for centre of mass and internal 
dynamics of an electromagnetically bound two-particle 
system by Sonnleitner and Barnett \cite{sonnleitner18} 
to the case including a weak post-Newtonian gravitational 
background field, described by the Eddington--Robertson 
parametrised post-Newtonian metric. Starting from a proper 
relativistic description of the situation, this approach 
allows to systematically derive the coupling of the model system 
to gravity, instead of `guessing' it by means of classical notions 
of relativistic effects.

We embed this technical result into a critical discussion 
concerning the problem of implementing and interpreting 
general couplings to the gravitational field and the connected 
problem of how to properly address the question concerning
the validity of the Equivalence Principle in Quantum Mechanics.
\end{abstract}

%%%%%%%%%%%%%%%%%%%%%%%%%%%%%%%%%%%%%%%%%%%%%%%%%%%%%
\newpage
\begin{footnotesize}
\setcounter{tocdepth}{3}
\tableofcontents
\end{footnotesize}

%%%%%%%%%%%%%%%%%%%%%%%%%%%%%%%%%%%%%%%%%%%%%%%%%%%%%%%%%%%
\section{Introduction}
\label{sec:Intro}
Experiments in quantum optics and matter-wave 
interferometry have now reached a degree of 
precision that covers `new' relativistic 
corrections that were hitherto not considered in such 
settings. In particular, this includes
couplings between `internal' and `centre of mass' 
degrees of freedom of composite systems, which 
have no Newtonian analogue.
Hence, such experiments 
require proper relativistic treatments for
their theoretical descriptions. Quite generally, such 
descriptions are often restricted to more or less 
\emph{ad hoc} addition of `effects' known from 
classical physics, like 
velocity-dependent masses, second-order Doppler 
shifts, second-order aberrations, redshifted 
energies, and  time dilations due to relative 
velocities and/or gravitational potentials. 
Such approaches are conceptually dangerous 
for a number of reasons. They neither guarantee 
completeness and independence, nor do they need 
to apply in non-classical situations where 
quantum properties dominate the dynamics. Such 
`derivations' often obscure the proper physical 
interpretation of relativistic corrections 
and, consequently, potentially 
misguide physical conclusions drawn form their 
observation. 

An example of this sort has recently been discussed 
in the careful and lucid analysis by Sonnleitner 
and Barnett \cite{sonnleitner18}, who pointed out 
that certain terms that show up in computations 
of the interaction of moving atoms with light, 
and which from a non-relativistic%
\footnote{As a matter of principle, we would like 
to avoid the common but misleading adjective 
`non-relativistic' to distinguish  Galilei 
invariant dynamical laws from `relativistic' 
ones, by which one then means those obeying 
Poincar\'e invariance. It is not the validity 
of the physical relativity principle that 
distinguishes both cases. Rather, their 
difference lies in the way that principle
is implemented. Since we cannot entirely 
escape traditionally established nomenclature 
without undue complications in expression, we 
will continue to use the term `non-relativistic' 
in the sense just explained and think of it 
as always being put between (invisible) 
inverted commas.} point of view 
appear to be some kind of `friction' (of a non 
immediately obvious physical origin, hence one 
may be tempted to call them `anomalous'), are 
actually nothing but a straightforward consequence 
of Special Relativity. In fact, there is nothing 
to be surprised about once the calculation is 
done, and the results are interpreted, in a proper
relativistic framework. This was done approximately 
in \cite{sonnleitner18} for the gravity-free case, 
by systematically deriving an `approximately 
relativistic' Hamiltonian describing the atom.
It is the purpose of our paper to extend this so as 
to also include gravity approximately. As emphasised 
above, this generalisation serves not only a point 
of principal interest that deserves clarification, 
but is also of immediate practical interest, not 
only in the obvious realm of quantum optics 
experiments regarding the detection of 
gravitational waves, but also in atom 
interferometry; see, 
e.g.~\cite{dimopoulos08,zych11,pikovski15,roura18,giese19,loriani19,zych19}. 
A calculation using methods very similar to those 
of \cite{sonnleitner18} including external 
gravitational fields was performed by Marzlin 
already in 1995 \cite{marzlin95}\footnote{We 
are grateful to Alexander Friedrich for 
pointing out this reference to us.}; but unlike 
Sonnleitner and Barnett in \cite{sonnleitner18} or 
our calculation in this work, Marzlin did not 
perform a full first-order post-Newtonian 
expansion and instead focused on the electric 
dipole coupling only.

What is generally needed is a consistent 
post-Newtonian approximation scheme in which the 
transition from Galilei invariant to 
Poincar\'e invariant (Special Relativity) and 
further to diffeomorphism invariant
(General Relativity) laws can be systematically 
derived in a step-by-step algorithmic fashion.
Such a scheme must be fundamentally rooted in 
a principle that contains all the information 
of how matter couples to gravity. 

For classical matter,
described in terms of dynamical laws compatible with 
the requirements of Special Relativity, such a 
principle is known in form of the `minimal coupling scheme'. 
According to this scheme, the Minkowski metric of
Special Relativity is to be replaced by the more 
general Lorentzian metric of spacetime and all partial 
derivatives (or more precisely: Levi--Civita covariant derivatives with respect to the Minkowski metric) are 
to be replaced with the Levi--Civita covariant 
derivatives with respect to the general spacetime 
metric. This scheme is based on Einstein's 
Equivalence Principle, whose core statement is 
that gravity can be encoded in the geometry 
of spacetime \emph{that is common to all matter 
components}. We stress that this is the important 
point encoding the \emph{universality} of 
gravitational interaction: that any matter component, 
may it be light, neutrinos, or other elementary 
particles with or without 
mass, spin, charge, or other features, or may 
it be a macroscopic body, like a football or a planet, 
all of them will couple to gravity in a way that 
only depends on one and the same geometry of 
spacetime; compare~\cite{thorne73}.

The problem we face in quantum physics as regards its 
coupling to gravitational fields is that the minimal 
coupling scheme has no straightforward generalisation 
to Galilei invariant Quantum Mechanics, for the obvious 
reason that the latter is incompatible with the 
requirements of Special Relativity. As is well known, 
enforcing Poincar\'e symmetry upon Quantum Mechanics 
eventually leads to the framework of Poincar\'e invariant Quantum 
Field Theory, often referred to as `Relativistic 
Quantum Field Theory' (RQFT), whose mathematical 
structure and physical interpretation is far more 
complex than that of ordinary non-relativistic
Quantum Mechanics. As a matter of principle, a 
post-Newtonian expansion should therefore start from 
RQFT.

However, if one is merely interested in 
leading order `relativistic corrections' below the 
threshold of pair production for any of the massive 
particles involved, a simpler method is to first put the 
\emph{classical} system in a fixed particle sector 
into Poincar\'e invariant form, then apply the minimal 
coupling scheme on the classical level, and finally 
apply suitable rules for quantisation, like the 
canonical ones. This is the procedure we will follow 
in this paper. It is computationally and conceptually 
much cheaper than a proper quantum field theoretic 
treatment and is \emph{a~priori} limited to sectors 
of fixed numbers in each massive particle species 
(hence it would not make sense to carry the computation 
of `relativistic corrections' further than to the 
lowest lying particle-production threshold).
In fact, the corrections we are after are far from those 
thresholds and may be safely derived from a systematic 
post-Newtonian expansion on the classical level. 

It is the underlying systematics of producing `relativistic corrections' that, in our opinion, distinguishes our 
approach from others, like, e.g., \cite{zych11,pikovski15,roura18,giese19,loriani19,zych19}. 
These latter attempts make use in an essential way of 
notions, like `wordline' and `redshift', which 
have no immediate meaning in quantum theory, unless 
the state of the system is severely restricted in an 
\emph{a priori} fashion. More precisely, the state 
of the total system (the atom) is assumed to 
unambiguously define a worldline, say for the 
centre of mass, which can also be assigned a length 
that is then identified with the proper time (up to 
division by a factor of $c$). This implies that 
1)~the overall pure state of the system separates 
into the tensor product of a pure state for the 
centre of mass with a pure state for the relative 
degrees of freedom, and that 
2)~the state for the centre of mass is of semiclassical 
nature, so as to unambiguously determine a piecewise 
smooth ($C^1$) worldline.%
\footnote{We recall that the path integral in ordinary 
Quantum Mechanics generally receives contributions 
from continuous but nowhere differentiable paths. 
Only in very special situations is the dominant 
contribution given by the action along a smooth 
classical path.}
It may well be that these 
\emph{a priori} restrictions can be justified in 
specific applications within quantum optics and atom interferometry. However, here we wish to promote the view 
that the theoretical problem of formulating a consistent 
post-Newtonian coupling scheme should be solved 
\emph{independently} of such restrictions.

In addition, phrasing relativistic corrections in 
terms of classical notions like `proper times', 
`redshifts', and the like bears the danger of 
losing control over aspects of completeness and 
possible redundancies. To be sure, these classical 
`effects' will appear as consequences from the general 
scheme, if applied to the specific situation that 
allows such vocabulary. But they should not be mistaken 
for the relativistic corrections proper, which derive 
from the spacetime geometry as a whole that enters 
the quantum dynamical laws at a fundamental level. 
Similar remarks apply in connection with attempts 
to formulate the Equivalence Principle in 
Quantum Mechanics. Since we consider this to be an important
and directly related issue, which also bears the 
danger of misconceptions, we will devote an extra 
section to its discussion at the end of this paper.

Problems of this sort were avoided by Sonnleitner and 
Barnett \cite{sonnleitner18} by the means of basing 
their whole calculation on a proper relativistic 
treatment of the situation (an atom interacting with 
an external electromagnetic field). In the end, the 
first-order post-Newtonian Hamiltonian they obtained 
could \emph{then} be used to interpret aspects of the 
situation in terms of classical `relativistic corrections':
The `centre of mass' part of the final Hamiltonian has the form 
of a single-particle kinetic Hamiltonian, where the 
r\^ole of the rest-mass of this particle is played by 
the total mass-energy of the atom, i.e. the sum of 
the rest masses of the consituent particles 
and the internal atomic energy divided by $c^2$. 
Thus, the computation in \cite{sonnleitner18}
explicitly shows that this physically intuitive 
picture of a `composite particle', suggested by 
mass--energy equivalence, can, in fact, be derived 
in a controlled and systematic approximation scheme, 
rather than merely made plausible from 
semi-intuitive physical considerations.

As will be shown in this paper,
a similar interpretation is possible for the 
situation including external 
gravitational fields: When expressing the final 
Hamiltonian using the physical spacetime metric, 
an intuitive `composite point particle' picture 
including the `mass defect' due to mass--energy 
equivalence will again be available for the 
centre of mass dynamics. Our essential result in 
that respect is expressed by equations 
\eqref{eq:Hamiltonian_com_C_final_full} to 
\eqref{eq:Hamiltonian_composite_point}.
This lends justification based 
on detailed calculations within systematic 
approximation schemes to some of the naiver 
approaches that are based on \emph{a priori} 
assumptions concerning the gravity--matter 
coupling.

%%%%%%%%%%%%%%%%%%%%%%%%%%%%%%%%%%%%%%%%%%%%%%%%%%%%%%%%%%%
\subsection{Logical structure of this paper}
\label{sec:Intro-structure}
In section \ref{sec:Situation-CS}, we set up the 
background for our calculations: after describing the 
physical system under consideration, we will give an overview over the 
method of computation in \cite{sonnleitner18}. Then we will 
discuss the geometric structures necessary to perform a 
post-Newtonian limit and to describe weak gravitational fields.

In the following, we will compute in detail the
`gravitational corrections' to the calculation by Sonnleitner and
Barnett \cite{sonnleitner18} arising from the presence of the
gravitational field. Section \ref{sec:Coupling-GF-Particle} will deal with the 
coupling of the gravitational field to the kinetic terms of the 
particles only, ignoring couplings of the gravitational to the electromagnetic field.

In section \ref{sec:Coupling-GF-EMF}, we will then compute the 
Lagrangian of the electromagnetic field in the presence 
of the gravitational field. This allows us to compute 
the total Hamiltonian describing the atomic system in section 
\ref{sec:Hamiltonian-Comp}, by repeating the calculation from section 
\ref{sec:Coupling-GF-Particle} while including the `gravitational 
corrections' to electromagnetism obtained in section~\ref{sec:Coupling-GF-EMF}. 
The resulting Hamiltonian will then be interpreted in terms of 
the physical spacetime metric and compared to earlier results 
in the remainder of section \ref{sec:Hamiltonian}.

In section \ref{sec:EP}, we will critically discuss 
attempts of formulating the Equivalence Principle in Quantum 
Mechanics, before concluding in section \ref{sec:Conclusion} 
with a brief summary of our paper.

In sections \ref{sec:Coupling-GF-Particle} and \ref{sec:Hamiltonian-Comp},
we will very closely follow the
calculation from and presentation in \cite{sonnleitner18}.
For the reader's convenience, we have reproduced all 
formulae from \cite{sonnleitner18} that are used in 
our calculation in the appendix.
We use the original numbering, prepended with
`\cite{sonnleitner18}.', so for example
\eqref{eq:Hamiltonian_com_cross_orig} refers to equation
(25f) of \cite{sonnleitner18}. As some of the equations 
from \cite{sonnleitner18} contain minor errors (mostly 
sign errors), we give the corrected versions in the 
appendix. The corresponding equation numbers are marked 
with a star, e.g. \eqref{eq:Hamiltonian_class_orig}.

%%%%%%%%%%%%%%%%%%%%%%%%%%%%%%%%%%%%%%%%%%%%%%%%%%%%%%%%%%%
\section{A composite system in external electromagnetic and gravitational fields}
\label{sec:Situation-CS}
We consider a simple system consisting of two 
particles without spin, with respective electric charges 
$e_1, e_2$, masses $m_1, m_2$, and spatial 
positions $\vect r_1, \vect r_2$. For simplicity 
we assume the charges to be equal and opposite, i.e.,
$e_2 = -e_1 =: e$. In what follows, we will take 
into account their mutual electromagnetic interaction 
but neglect their mutual gravitational interaction. 
This two-particle system, which we will sometimes 
refer to as `atom', will be placed in an external electromagnetic field, which we will take into 
account, as well as an external gravitational 
field, that we will also take into account. It is 
the inclusion of the latter that we wish to discuss
in this paper and that extends previous 
studies \cite{sonnleitner18}.

%%%%%%%%%%%%%%%%%%%%%%%%%%%%%%%%%%%%%%%%%%%%%%%%%%%%%%%%%%%
\subsection{External electromagnetic fields -- the work of Sonnleitner and Barnett}
\label{sec:Situation-CS-EM}
In \cite{sonnleitner18}, Sonnleitner and Barnett
describe a systematic method to obtain an 
`approximately relativistic' quantum Hamiltonian 
for a system as described above interacting with 
an external electromagnetic field, where `approximately 
relativistic' refers to the inclusion of lowest 
order post-Newtonian correction terms, i.e. of 
order $c^{-2}$. Their work was motivated by their 
own observation \cite{sonnleitner17,sonnleitner18a} 
that the electromagnetic interaction of a decaying 
atom, which in QED follows an intrinsically 
special-relativistic symmetry (i.e. Poincar\'e 
invariance), will give rise to unnaturally looking 
friction-like terms that seem to contradict the 
relativity principle (which, of course, they don't)
if interpreted in a non-relativistic (i.e. Galilei invariant) setting of ordinary Quantum Mechanics. 
Their correct conclusion in \cite{sonnleitner18}
was that this confusion can be altogether avoided 
by replacing this `hotchpotch' (their wording, see
last line on p. 042106-9 of \cite{sonnleitner18}) 
of symmetry concepts by a systematic post-Newtonian derivation starting from a common, manifestly 
Poincar\'e symmetric description.

As our development will closely follow theirs, let us 
describe the strategy of \cite{sonnleitner18} in 
more detail. They start with the classical 
Poincar\'e invariant Lagrangian function describing 
the situation, where the kinetic terms of the 
two particles are expanded to post-Newtonian order. 
This Lagrangian includes the electromagnetic fields 
generated by the particles themselves as lowest-order
solutions of the Maxwell equations, thus eliminating 
the `internal' field degrees of freedom.

This classical Lagrangian (the sum of the famous Darwin
Lagrangian \cite{darwin20} and the external field 
Lagrangian) is then Legendre transformed and canonically
quantised to obtain a quantum Hamiltonian in what they 
call the `minimal coupling form'. They then perform a 
Power--Zienau--Woolley (PZW) unitary transformation
\cite{power59, woolley71, babiker83} together with a
multipolar expansion of the external field in order 
to transform the Hamiltonian into a so-called 
`multipolar form'.

Then, introducing Newtonian centre of mass and relative coordinates $\vect R, \vect r$, and the corresponding 
canonical momenta $\vect P, \vect p_{\vect r}$, they 
arrive at what they call the centre of mass Hamiltonian.
This can be separated into parts describing the 
central motion of the atom, the internal atomic 
motion, the external electromagnetic field, and its 
interaction with the atom. However, this 
Hamiltonian also contains 
cross terms coupling the relative degrees of freedom 
to the central momentum $\vect P$. In order to 
eliminate this coupling, they perform a final 
canonical transformation to new coordinates 
$\vect Q, \vect q$ and momenta $\vect P, \vect p$.

%%%%%%%%%%%%%%%%%%%%%%%%%%%%%%%%%%%%%%%%%%%%%%%%%%%%%%%%%%%
\subsection{Including weak external gravitational fields}
\label{sec:Situation-CS-GF}
As already stated above, our contribution in this 
paper will consist in generalising the calculation 
of \cite{sonnleitner18} to the case of the atom 
being situated in a weak external gravitational 
field in addition to the electromagnetic field already 
considered in  \cite{sonnleitner18}. Our aim is to 
likewise obtain an `approximately relativistic', 
i.e. first-order post-Newtonian, Hamiltonian 
describing this situation.

Conceptually speaking, this generalisation is not
entirely obvious for the following reason: The 
addition of gravitational fields will, according 
to General Relativity, result in a changed geometry 
of spacetime and, consequently, in the loss of some 
or all spacetime symmetries and their associated 
conservation laws. In particular, Poincar\'e 
symmetry will be lost and there is no obvious way 
to implement the post-Newtonian expansion employed 
in \cite{sonnleitner18}. In fact, the concept of a 
`post-Newtonian expansion' in an arbitrary spacetime 
simply does not exist without the explicit 
introduction of certain background structures that 
give meaning to notions like `weak' gravitational 
fields and `slow' velocities. To cut a long story 
short, such necessary background structures will 
in our case be 
1)~the Poincar\'e symmetric Minkowski metric 
$\eta$ on spacetime $M$ and 
2)~a preferred inertial reference frame in 
Minkowski space $(M,\eta)$, mathematically 
represented by a timelike vector field $u$ which is 
geodesic for $\eta$.

That the gravitational field be weak then 
means that the physical spacetime metric $g$ 
deviates only little from the Minkowski metric 
$\eta$. More precisely this means in our case 
that quadratic and higher orders of the difference 
$h:=g-\eta$ and its derivatives can be neglected. 
That velocities be slow means that the velocity
$v$ of each particle relative to the preferred 
frame is small to $c$, so that terms in $v/c$
of higher order than the second can be neglected.

We stress that all the structures introduced 
and all the conditions of `weakness' and 
`slowness' mentioned are totally independent 
of coordinates that we may choose to parametrise 
spacetime. That is not to say that there may 
not be preferred coordinates which are 
particularly adapted to the given background 
structure. Indeed, such adapted coordinates 
obviously exist, namely so-called inertial 
coordinates $\{x^0,x^1,x^2,x^3\}$ in Minkowski 
space $(M,\eta)$, such that $x^0 = ct$, 
$u = \partial/\partial t$, and $\eta = \eta_{\mu\nu} \, \D x^\mu \otimes \D x^\nu$ with $(\eta_{\mu\nu}) = \mathrm{diag}(-1,1,1,1)$. In the same coordinate system, 
the components of the physical spacetime 
metric $g = g_{\mu\nu} \, \D x^\mu \otimes \D x^\nu$ that we 
consider here are then given by
\begin{equation} \label{eq:metric}
	(g_{\mu\nu}) = \begin{pmatrix}
		-1 - 2 \frac{\phi}{c^2} - 2 \beta \frac{\phi^2}{c^4} + \Or(c^{-6}) & \Or(c^{-5})\\
		\Or(c^{-5}) & (1 - 2 \gamma \frac{\phi}{c^2})\mathds{1} + \Or(c^{-4})
	\end{pmatrix}
\end{equation}
where $\phi$ is a scalar function on spacetime that 
may be seen as the analogue of the Newtonian gravitational 
potential in this approximation scheme. 
The weakness of the gravitational field is expressed in components by $|h_{\mu\nu}| = |g_{\mu\nu}-\eta_{\mu\nu}| \ll 1$. Using 
\eqref{eq:metric} this is equivalent to the smallness
of the Newtonian potential as compared to $c^2$, that is 
$\phi/c^2\ll 1$.

The metric \eqref{eq:metric} also contains two 
dimensionless parameters $\beta$ and $\gamma$,
the so-called `Eddington--Robertson parameters', 
which we introduced in order to account for 
possible deviations from General Relativity. General
Relativity corresponds to the values $\beta=\gamma=1$, 
in which case the metric \eqref{eq:metric} solves 
the field equations of General Relativity 
approximately in a $1/c$-expansion for a static 
source, with $\phi$ being the Newtonian gravitational 
potential of the source. The metrics for different values of these 
parameters are then considered to correspond to 
so-called `test theories' against which the 
predictions of General Relativity can be tested; 
see, e.g.,~\cite{will93}. Following standard 
terminology, we shall refer to \eqref{eq:metric} 
as the `PPN metric' (parametrised post-Newtonian).
The explicit inclusion of $\beta$ and $\gamma$ in our formalism allows us 
to track the consequences of post-Newtonian 
corrections in the spatial and the temporal part of the 
metric separately. This also opens the possibility 
to apply our results to potential future 
quantum tests of General Relativity itself, which 
are outside the scope of this paper.

For later use, we introduce the `physical spatial metric' 
$^{(3)}g$, which is the restriction of the physical 
spacetime metric $g$ to three-dimensional `space', i.e. to 
the orthogonal complement of the preferred vector field $u$. 
The inverse of this physical spatial metric will be denoted 
by $^{(3)}g^{-1}$.

Later we will also need the inverse metric to $g$, 
the components of which in the specified coordinate 
system are simply obtained by inverting the matrix 
\eqref{eq:metric}:
\begin{equation} \label{eq:metric_inv}
	(g^{\mu\nu}) = \begin{pmatrix}
		-1 + 2 \frac{\phi}{c^2} + (2\beta - 4) \frac{\phi^2}{c^4} + \Or(c^{-6}) & \Or(c^{-5})\\
		\Or(c^{-5}) & (1 + 2 \gamma \frac{\phi}{c^2})\mathds{1} + \Or(c^{-4})
	\end{pmatrix}
\end{equation}

Since we are interested in a lowest-order post-Newtonian description, we will work up to (and including) terms of order $c^{-2}$ and neglect higher order terms. In fact, corrections of higher order cannot be treated in a simple Hamiltonian formalism as employed here, without explicitly including the internal electromagnetic field degrees of freedom as dynamical variables: Elimination of the internal field variables by solving Maxwell's equations will introduce retardation effects at higher orders, thus leading to an action that is non-local in time, spoiling the application 
of the Hamiltonian formalism. 

%%%%%%%%%%%%%%%%%%%%%%%%%%%%%%%%%%%%%%%%%%%%%%%%%%%%%%%%%%%
\subsection{Geometric structures, notation, and conventions} 
\label{sec:Situation-CS-Conventions}
The chosen background structures 
that allow us to define the approximation scheme 
(according to `slow' and `weak') are also employed in 
developing our calculation in parallel with that in \cite{sonnleitner18}.
More precisely, we use the background Minkowski 
metric $\eta$ and the preferred timelike vector field 
$u$ to decompose spacetime into time (integral lines 
of $u$)  and space (hyperplanes $\eta$-perpendicular
to $u$), and to endow space with a flat Riemannian 
metric (the restriction of $\eta$ to the hyperplanes).
With that structure `space' just becomes ordinary 
flat Euclidean space. We are indeed free to use this 
`flat' structure to perform all our computations, 
the benefit being the aimed-for direct comparison 
with \cite{sonnleitner18}. However, once the results 
of the computations are established, we have to 
keep in mind that physical distances and times are 
measured with the physical metric $g$, not the 
auxiliary metric $\eta$. We will see that it is 
precisely this re-interpretation of the formulae 
obtained that lends them good physical meaning. 

In our calculations, vectors and tensors will be 
represented by their components with respect to 
the chosen coordinate system $(x^\mu) = (ct, x^a)$.
We let greek indices run from $0$ to $3$, latin indices
from $1$ to $3$ and we shall use the Einstein summation 
convention for like indices at different levels (one 
up- one downstairs). 
Indices are lowered and raised by the physical spacetime 
metric $g_{\mu\nu}$ and its inverse $g^{\mu\nu}$ respectively. 
The Minkowski metric takes its 
usual diagonal form (as stated above) and the spatial 
metric its usual Euclidean form with components 
of the metric tensor given by 
$(\delta_{ab}) = \mathrm{diag}(1,1,1)$ and its 
inverse $(\delta^{ab}) = \mathrm{diag}(1,1,1)$. 

We will often employ a `three-vector' notation, where 
the three-tuple of spatial components of some quantity 
will be denoted by an upright, boldface letter: 
for example, $\vect r_1$ is the `vector' of spatial
 coordinates of the first particle. When using this 
notation, a dot between two such `vectors' will 
denote the \emph{component-wise} `Euclidean scalar 
product', i.e.
\begin{equation} \label{eq:EuclVectorprod}
	\vect v \cdot \vect w := \delta_{ab} v^a w^b = \sum_{a=1}^3 v^a w^a,
\end{equation}
and a cross multiplication symbol will denote the
\emph{component-wise} vector product, i.e.
\begin{equation}
	(\vect v \times \vect w)^a := \delta^{an} \varepsilon_{nbc} v^b w^c
\end{equation}
where $\varepsilon_{abc}$ is the usual totally antisymmetric 
symbol. Geometrically, $\varepsilon_{abc}$ can be understood 
as the components of the spatial volume form induced by the 
Euclidean metric. We will lower and raise the indices of 
$\varepsilon$ by $\delta_{ab}$ and $\delta^{ab}$ respectively,
i.e. $\varepsilon^a_{\phantom{a}bc}:=\delta^{an}\varepsilon_{nbc}$ etc.,
such that we can write 
$(\vect v \times \vect w)^a = \varepsilon^a_{\phantom{a}bc} v^b w^c$.

A boldface nabla symbol $\vect\nabla$ denotes the 
three-tuple of \emph{partial} derivatives,
\begin{equation} \label{eq:Nabla}
	\vect \nabla = (\partial_1, \partial_2, \partial_3),
\end{equation}
which can be geometrically understood as the component 
representation of the spatial covariant derivatives 
with respect to the flat euclidean metric. It will 
be used to express component-wise vector calculus 
operations in the usual short-hand notation, 
for example writing
\begin{equation}
	(\vect \nabla \times \vect A)^a = \varepsilon^{abc} \partial_b A_c
\end{equation}
for the component-wise curl of $\vect A$. 

In view of the structures introduced we stress again 
that all the operations reported here and used in the 
sequel make good geometric sense. They do depend on 
the geometric structures that we made explicit above, 
but they do not depend on the coordinates or frames 
that one uses in order to express the geometric 
objects (including the background structures) in 
terms of their real-valued components. This fact 
is very important to keep in mind if it comes to 
the task of interpreting the results of computations.
For example, these results will contain geometric 
operations, like scalar products, which my be taken 
using either of the metric structures provided by the
formalism. What may then appear as a more or less 
complicated gravitational correction to the flat 
space result will then, in fact, turn out to be a 
simple and straightforward transcription of the latter into 
the proper physical metric, as one might have 
anticipated from some more or less 
naive working-version of the Equivalence Principle. 
Interpretational issues like this are well known in 
the literature on gravitational couplings of quantum
systems; see, e.g.,  \cite{marzlin95,laemmerzahl95}.
For us, too, they will once more turn out to be 
relevant in connection with the total Hamiltonian 
in section~\ref{sec:Hamiltonian-Comp} and the 
ensuing discussion in 
section~\ref{sec:Hamiltonian-EM-ON}. We will
derive and interpret the relevant gravitational 
terms relative to the background structures 
$(\eta,u)$ in order to keep the analogy with the 
computation in \cite{sonnleitner18}, but then we 
shall re-interpret the results in terms of the 
proper physical metric $g$ in order to reveal 
their naturalness.

%%%%%%%%%%%%%%%%%%%%%%%%%%%%%%%%%%%%%%%%%%%%%%%%%%%%%%%%%%%
\section{Coupling the gravitational field to the particles}
\label{sec:Coupling-GF-Particle}
In this section we will work out the influence of the gravitational field when coupled to the kinetic terms of the particles only, ignoring its couplings to the electromagnetic field. The latter will be the subject of the following sections.

Starting from the Lagrangian for our atom in the \emph{absence} of gravity and adding the `gravitational corrections' to the kinetic terms of the particles, we will then repeat the calculation of \cite{sonnleitner18} to obtain a quantum Hamiltonian in centre of mass coordinates.

%%%%%%%%%%%%%%%%%%%%%%%%%%%%%%%%%%%%%%%%%%%%%%%%%%%%%%%%%%%
\subsection{The classical Hamiltonian}
\label{sec:Class-Ham}
For a single free point particle with mass $m$ and position $\vect x$, the classical kinetic Lagrangian (parametrising the worldline by coordinate time) in our metric \eqref{eq:metric} reads
\begin{align}\label{eq:Lagrangian_point_grav}
	L_\text{point} &= -mc^2 \sqrt{-g_{\mu\nu} \dot x^\mu \dot x^\nu/c^2} \nonumber\\
	&= \frac{m \dot{\vect x}^2}{2} \left(1 + \frac{\dot{\vect x}^2}{4 c^2}\right) - m c^2 - m\phi \left(1 + (2\beta-1) \frac{\phi}{2c^2}\right) - \frac{2\gamma + 1}{2} \frac{m\phi}{c^2} \dot{\vect x}^2 + \Or(c^{-4}).
\end{align}
Now considering our two-particle system, the kinetic terms for the particles in gravity are given as the sum of two terms as in \eqref{eq:Lagrangian_point_grav}. These lowest-order `gravitationally corrected' kinetic terms we include into the classical Lagrangian from \eqref{eq:Lagrangian_class_start_orig}\footnote
	{We remind the reader that all the equations from \cite{sonnleitner18} that we refer to explicitly are reproduced in the appendix.},
which described two particles interacting with an electromagnetic field in the \emph{absence} of gravity.

Eliminating the internal electromagnetic fields literally as in \cite{sonnleitner18}, we arrive at the post-Newtonian classical Lagrangian
\begin{align}
	L_\text{new} &= L - m_1 \phi(\vect r_1) - m_2 \phi(\vect r_2) - \frac{2\gamma + 1}{2} \frac{m_1\phi(\vect r_1)}{c^2} \dot{\vect r}_1^2 - \frac{2\gamma + 1}{2} \frac{m_2\phi(\vect r_2)}{c^2} \dot{\vect r}_2^2 \nonumber
		\\&\quad - (2\beta-1) \frac{m_1 \phi(\vect r_1)^2}{2c^2} - (2\beta-1) \frac{m_2 \phi(\vect r_2)^2}{2c^2}
\end{align}
describing our electromagnetically bound two-particle system and the external electromagnetic field. Here $L$ is the final classical Lagrangian from \eqref{eq:Lagrangian_class_postNewt_orig}, \eqref{eq:Lagrangian_class_Darwin_orig}.

Legendre transforming this with respect to the particle velocities $\dot{\vect r}_i$ and the time derivative $\partial_t \vect A^\perp$ of the electromagnetic vector potential, we obtain the total classical Hamiltonian
\begin{align} \label{eq:Hamiltonian_class_no_em}
	H_\text{new} &= H + m_1 \phi(\vect r_1) + m_2 \phi(\vect r_2) + \frac{2\gamma+1}{2 m_1 c^2} \phi(\vect r_1) \bar{\vect p}_1^2 + \frac{2\gamma+1}{2 m_2 c^2} \phi(\vect r_2) \bar{\vect p}_2^2 \nonumber
		\\&\quad+ (2\beta-1) \frac{m_1 \phi(\vect r_1)^2}{2c^2} + (2\beta-1) \frac{m_2 \phi(\vect r_2)^2}{2c^2}.
\end{align}
Here $H$ is the classical Hamiltonian from \eqref{eq:Hamiltonian_class_orig} and $\bar{\vect p}_i = \vect p_i - e_i \vect A^\perp(\vect r_i)$ is the kinetic momentum. Note that we dropped all terms that go beyond our order of approximation.

%%%%%%%%%%%%%%%%%%%%%%%%%%%%%%%%%%%%%%%%%%%%%%%%%%%%%%%%%%%
\subsection{Canonical quantisation and PZW transformation to a multipolar Hamiltonian}
Now, we canonically quantise this Hamiltonian and perform the PZW transformation and electric dipole approximation used in \cite{sonnleitner18} to arrive at the `multipolar' Hamiltonian from \eqref{eq:Hamiltonian_mult_orig}.
Neglecting terms of the form $\frac{\vect p_i \cdot [\vect d \times \vect B(\vect R)]}{m_i m_j c^2}$ as in \eqref{eq:SB_negl_interaction}, in our gravitational correction terms from \eqref{eq:Hamiltonian_class_no_em} these transformations amount just to the replacement $\bar{\vect p}_i \to \vect p_i$ (compare the appendix from \eqref{eq:PZW} to \eqref{eq:SB_21}). Hence the multipolar Hamiltonian including the gravitational correction terms is
\begin{align} \label{eq:Hamiltonian_mult_no_em}
	H_\text{[mult],new} &= H_\text{[mult]} + m_1 \phi(\vect r_1) + m_2 \phi(\vect r_2) + \frac{2\gamma+1}{2 m_1 c^2} \vect p_1 \cdot \phi(\vect r_1) \vect p_1 + \frac{2\gamma+1}{2 m_2 c^2} \vect p_2 \cdot \phi(\vect r_2) \vect p_2 \nonumber
		\\&\quad+ (2\beta-1) \frac{m_1 \phi(\vect r_1)^2}{2c^2} + (2\beta-1) \frac{m_2 \phi(\vect r_2)^2}{2c^2},
\end{align}
where $H_\text{[mult]}$ is the multipolar Hamiltonian from \eqref{eq:Hamiltonian_mult_orig}.

Now that we are on the quantum level, we had to choose a symmetrised operator ordering for the $\vect p^2\phi$ terms. We chose an ordering of the form $\vect p \cdot \phi \vect p$ since that is an `obvious' choice, which also results if a WKB-like expansion of the Klein--Gordon equation is used for the description of single quantum particles in an Eddington--Robertson PPN metric \cite{schwartz19}, when one neglects terms involving $\Delta\phi$ (which vanishes outside the matter generating the Newtonian potential $\phi$).

%%%%%%%%%%%%%%%%%%%%%%%%%%%%%%%%%%%%%%%%%%%%%%%%%%%%%%%%%%%
\subsection{Introduction of centre of mass variables}
We now want to express the correction terms in (Newtonian) centre of mass and relative variables,
\begin{align}
	\vect R &= \frac{m_1 \vect r_1 + m_2 \vect r_2}{M} \; , & \vect r &= \vect r_1 - \vect r_2 \; ,\\
	\vect P &= \vect p_1 + \vect p_2 \; , & \vect p_{1,2} &= \frac{m_{1,2}}{M} \vect P \pm \vect p_{\vect r} \; ,
\end{align}
where $M = m_1 + m_2$. To this end, we expand the gravitational potential $\phi$ around the centre of mass position $\vect R$ \emph{in linear order} (i.e. perform a monopole approximation of the generating mass distribution). In this approximation, we have $m_1 \phi(\vect r_1) + m_2 \phi(\vect r_2) = M \phi(\vect R)$ and $m_1 \phi(\vect r_1)^2 + m_2 \phi(\vect r_2)^2 = M \phi(\vect R)^2$. Furthermore using
\begin{align}
	\vect p_{1,2} \cdot \phi(\vect r_{1,2}) \vect p_{1,2} &= \left(\frac{m_{1,2}}{M} \vect P \pm \vect p_{\vect r}\right) \cdot \phi(\vect r_{1,2}) \left(\frac{m_{1,2}}{M} \vect P \pm \vect p_{\vect r}\right) \nonumber \\
	&= \frac{m_{1,2}^2}{M^2} \vect P \cdot \phi(\vect r_{1,2}) \vect P \pm \frac{m_{1,2}}{M} (\vect P \cdot \phi(\vect r_{1,2}) \vect p_{\vect r} + \text{H.c.}) + \vect p_{\vect r} \cdot \phi(\vect r_{1,2}) \vect p_{\vect r}
\end{align}
and the relations $\phi(\vect r_1) - \phi(\vect r_2) = \vect r \cdot \vect\nabla\phi(\vect R)$ as well as
\begin{align}
	\frac{1}{m_1} \phi(\vect r_1) + \frac{1}{m_2} \phi(\vect r_2) &= \left(\frac{1}{m_1} + \frac{1}{m_2}\right) \phi(\vect R) + \frac{1}{M} \left(\frac{m_2}{m_1} - \frac{m_1}{m_2}\right) \vect r \cdot \vect\nabla\phi(\vect R) \nonumber \\
	&= \frac{1}{\mu} \phi(\vect R) - \frac{m_1 - m_2}{m_1 m_2} \vect r \cdot \vect\nabla\phi(\vect R)
\end{align}
where $\mu = \frac{m_1 m_2}{M}$ is the system's reduced mass, we arrive at the centre of mass Hamiltonian
\begin{align} \label{eq:Hamiltonian_com_no_em}
	H_\text{[com],new} &= H_\text{[com]} + M \phi(\vect R) + (2\beta-1) \frac{M \phi(\vect R)^2}{2 c^2} + \frac{2\gamma+1}{2Mc^2} \vect P \cdot \phi(\vect R) \vect P \nonumber
		\\&\quad+ \frac{2\gamma+1}{2\mu c^2} \vect p_{\vect r}^2 \phi(\vect R) + \frac{2\gamma+1}{2Mc^2} \left[\vect P \cdot (\vect r \cdot \vect\nabla\phi(\vect R)) \vect p_{\vect r} + \text{H.c.}\right] \nonumber
		\\&\quad- \frac{2\gamma+1}{2c^2} \frac{m_1 - m_2}{m_1 m_2} \vect p_{\vect r} \cdot (\vect r \cdot \vect\nabla\phi(\vect R)) \vect p_{\vect r} \; ,
\end{align}
where $H_\text{[com]}$ is the centre of mass Hamiltonian from \eqref{eq:Hamiltonian_com_orig}.

This can, as in \cite{sonnleitner18}, be brought into the form
\begin{equation}
	H_\text{[com],new} = H_\text{C,new} + H_\text{A,new} + H_\text{AL} + H_\text{L} + H_\text{X,new},
\end{equation}
where
\begin{equation}
	H_\text{C,new} = H_\text{C} + \frac{2\gamma+1}{2Mc^2} \vect P \cdot \phi(\vect R) \vect P + \left(M + \frac{\vect p_{\vect r}^2}{2\mu c^2}\right) \phi(\vect R) + (2\beta-1) \frac{M \phi(\vect R)^2}{2 c^2}
\end{equation}
describes the dynamics of the centre of mass and
\begin{equation} \label{eq:Hamiltonian_com_atom_no_em}
	H_\text{A,new} = H_\text{A} + 2\gamma\frac{\phi(\vect R)}{c^2} \frac{\vect p_{\vect r}^2}{2\mu} - \frac{2\gamma+1}{2c^2} \frac{m_1 - m_2}{m_1 m_2} \vect p_{\vect r} \cdot (\vect r \cdot \vect\nabla\phi(\vect R)) \vect p_{\vect r}
\end{equation}
describes the internal dynamics of the atom, both modified in comparison to \cite{sonnleitner18}. Here, we have included the term $2\gamma\frac{\phi(\vect R)}{c^2} \frac{\vect p_{\vect r}^2}{2\mu}$ into $H_\text{A,new}$ since it can be combined with $\frac{\vect p_{\vect r}^2}{2\mu}$ from $H_\text{A}$ into
\begin{equation} \label{eq:internal_kin_metr}
	\frac{\vect p_{\vect r}^2}{2\mu} \left(1 + 2\gamma\frac{\phi(\vect R)}{c^2}\right) = \frac{{^{(3)}g^{-1}_{\vect R}} (\vect p_{\vect r}, \vect p_{\vect r})}{2\mu},
\end{equation}
giving the geometrically correctly defined Newtonian internal kinetic energy, using the metric square of the internal momentum. Here $^{(3)}g_{\vect R}^{-1}$ denotes the inverse of the physical spatial metric at position $\vect R$, as explained before \eqref{eq:metric_inv}.

The terms $H_\text{AL}$ and $H_\text{L}$ describing, respectively, the atom-light interaction and the electromagnetic field are not changed compared to \cite{sonnleitner18}; and the final summand
\begin{equation}
	H_\text{X,new} = H_\text{X} + \frac{2\gamma+1}{2Mc^2} \left[\vect P \cdot (\vect r \cdot \vect\nabla\phi(\vect R)) \vect p_{\vect r} + \text{H.c.}\right]
\end{equation}
containing `cross terms' coupling the internal degrees of freedom to the central momentum $\vect P$ gains an additional term.

Note that if we assume that the gravitational potential $\phi$ vary slowly over the extension of the atom, we can neglect the terms $\vect r \cdot \vect\nabla\phi(\vect R)$.

%%%%%%%%%%%%%%%%%%%%%%%%%%%%%%%%%%%%%%%%%%%%%%%%%%%%%%%%%%%
\section{Coupling the gravitational to the electromagnetic field}
\label{sec:Coupling-GF-EMF}

Having determined the gravitational field's coupling to the particles in the previous section, we now turn to its coupling to the electromagnetic field. In the following section \ref{sec:Hamiltonian} we will then combine all couplings into a single Hamiltonian.

%%%%%%%%%%%%%%%%%%%%%%%%%%%%%%%%%%%%%%%%%%%%%%%%%%%%%%%%%%%
\subsection{Solution of the gravitationally modified Maxwell equations}
The electromagnetic part of the action, including interaction with matter, is
\begin{equation} \label{eq:em_action}
	S_\text{em} = \int\D t\D^3\vect x \, \sqrt{-g} \left(-\frac{1}{4\mu_0} F_{\text{tot.}\mu\nu}F_\text{tot.}^{\mu\nu} + J^\mu A_{\text{tot.}\mu}\right),
\end{equation}
where $g$ denotes the determinant of the matrix $(g_{\mu\nu})$ of components of the metric, $J = J^\mu \partial_\mu$ is the 4-current `density' vector field, $A_\text{tot.} = A_{\text{tot.}\mu} \D x^\mu$ is the (total\footnote{We will later decompose the field into internal and external contributions, hence the label `total'.}) electromagnetic 4-potential form and $\D A_\text{tot.} = F_\text{tot.} = F_{\text{tot.}\mu\nu} \D x^\mu \otimes\D x^\nu = (\partial_\mu A_{\text{tot.}\nu} - \partial_\nu A_{\text{tot.}\mu}) \D x^\mu \otimes\D x^\nu$ is the electromagnetic field tensor.
This is the standard action describing electromagnetism in a gravitational field, which is obtained by minimally coupling the special-relativistic action for electromagnetism \cite{jackson98} to a general spacetime metric \cite{misner73,hawking73}.

Note that $J^\mu$ are the components of a proper vector field and not of a density; their relation to the 4-current \emph{density} with components $j^\mu$, in terms of which the interaction part of the action takes the form $\int\D t\D^3\vect x \, j^\mu A_{\text{tot.}\mu}$, is given by
\begin{equation}
	J^\mu = \frac{1}{\sqrt{-g}} j^\mu.
\end{equation}
The current density of our system of two particles is given by\footnote
	{For a single particle of charge $q$ on an arbitrarily parametrised timelike worldline $r^\mu(\lambda)$, the current density is given by
	\[j^\mu(x) = q c\int\D\lambda \, \frac{\D r^\mu}{\D\lambda} \delta^{(4)}\bigl(x - r(\lambda)\bigr).\]
	Parametrising by coordinate time and considering two particles, we arrive at the above expression.}
\begin{equation} \label{eq:current_dens}
	j^\mu(t,\vect x) = \sum_{i=1}^2 e_i \delta^{(3)}(\vect x - \vect r_i(t)) \dot r_i^\mu(t),
\end{equation}
where the dot denotes differentiation with respect to coordinate time $t$. The charge density is
\begin{equation} \label{eq:charge_dens}
	\rho := \frac{1}{c} j^0.
\end{equation}
The Maxwell equations obtained by varying the action with respect to $A_{\text{tot.}\mu}$ take the form
\begin{equation}
	\nabla_\mu F_\text{tot.}^{\mu\nu} = - \mu_0 J^\nu
\end{equation}
in terms of the current vector field, or
\begin{equation}
	\nabla_\mu F_\text{tot.}^{\mu\nu} = - \mu_0 \frac{1}{\sqrt{-g}} j^\nu
\end{equation}
in terms of the current density. It will be useful to consider the form
\begin{equation} \label{eq:Maxwell}
	\nabla^\mu F_{\text{tot.}\mu\nu} = - \mu_0 \frac{1}{\sqrt{-g}} j_\nu
\end{equation}
instead.

\emph{From now on we employ the approximation that over the extension of the atom, the gravitational field $\phi$ is constant.} This implies that all partial derivatives of the components of the metric vanish, such that the Christoffel symbols vanish and all covariant derivatives are given just by partial derivatives. Furthermore, we employ the Coulomb gauge $\nabla^i A_{\text{tot.}i} = 0$. Then, the Maxwell equations \eqref{eq:Maxwell} become
\begin{equation} \label{eq:Maxwell_simpl_tot}
	- \mu_0 \frac{1}{\sqrt{-g}} j_\nu = g^{\mu\rho} \partial_\mu \partial_\rho A_{\text{tot.}\nu} - g^{0\rho} \partial_\nu \partial_\rho A_{\text{tot.}0} \; .
\end{equation}

We now split the electromagnetic potential into `internal' and `external' parts, as done in \cite{sonnleitner18}, i.e. the part generated by our system of moving particles and that corresponding to external electromagnetic fields: We have $A_{\text{tot.}\mu} = \mathcal A_\mu + A_\mu$, where we adopt the Coulomb gauge for both the external part $A_\mu$ and the internal part $\mathcal A_\mu$. The external potential $A_\mu$ is assumed as given and satisfying the vacuum Maxwell equations, and the internal potential satisfies the Maxwell equations for the internal current density \eqref{eq:current_dens}, i.e.
\begin{equation} \label{eq:Maxwell_simpl}
	- \mu_0 \frac{1}{\sqrt{-g}} j_\nu = g^{\mu\rho} \partial_\mu \partial_\rho \mathcal A_\nu - g^{0\rho} \partial_\nu \partial_\rho \mathcal A_0 \; .
\end{equation}
Similarly, we write $F_{\text{tot.}\mu\nu} = \mathcal F_{\mu\nu} + F_{\mu\nu}$, where $\mathcal F = \D\mathcal A$ is the internal and $F = \D A$ is the external field tensor.

Inserting the PPN metric \eqref{eq:metric_inv}, the $0$-component of the Maxwell equations \eqref{eq:Maxwell_simpl} reads as follows:
\begin{align} \label{eq:Maxwell_zero_comp}
	- \mu_0 \frac{1}{\sqrt{-g}} j_0 &= g^{\mu\rho} \partial_\mu \partial_\rho \mathcal A_0 - g^{0\rho} \partial_0 \partial_\rho \mathcal A_0 \nonumber \\
	&= g^{a\rho} \partial_a \partial_\rho \mathcal A_0 \nonumber \\
	&= c^{-1} g^{0a} \partial_t \partial_a \mathcal A_0 + g^{ab} \partial_a \partial_b \mathcal A_0 \nonumber \\
	&= \left(1 + 2\gamma \frac{\phi}{c^2}\right) \Delta \mathcal A_0 + \Or(c^{-4})
\end{align}
Here $\Delta = \delta^{ab} \partial_a \partial_b$ denotes the `flat' spatial Laplace operator with respect to the spatial Euclidean metric defined by the background structures (flat Minkowski metric and preferred time-like vector field).

Now, using\footnote
	{The determinant of the spatial metric is $\det(g_{ab}) =: {^{(3)}g} = 1 - 6 \gamma \frac{\phi}{c^2} + \Or(c^{-4})$, and so minus the determinant of the metric is $-g = {^{(3)}g} / (-g^{00}) = (1 - 6 \gamma \frac{\phi}{c^2} + \Or(c^{-4})) \cdot (1 + 2 \frac{\phi}{c^2} + \Or(c^{-4})) = 1 - 2 (3\gamma - 1) \frac{\phi}{c^2} + \Or(c^{-4})$, implying $\sqrt{-g} = 1 - (3\gamma - 1) \frac{\phi}{c^2} + \Or(c^{-4})$.
	}
$1/\sqrt{-g} = 1 + (3\gamma-1) \frac{\phi}{c^2} + \Or(c^{-4})$, this equation is equivalent to
\begin{equation}
	\Delta \mathcal A_0 = - \left(1 + (\gamma-1) \frac{\phi}{c^2}\right) \mu_0 j_0 + \Or(c^{-4}).
\end{equation}
Rewriting $j_0 = -(1 + 2\frac{\phi}{c^2}) j^0 + \Or(c^{-4})$, recalling the charge density \eqref{eq:charge_dens} and introducing the internal electric scalar potential
\begin{equation}
	\phi_\text{el.} := - c \mathcal A_0,
\end{equation}
we arrive at the Poisson equation for the electric potential:
\begin{equation} \label{eq:Poisson_el_pot}
	\Delta \phi_\text{el.} = - \frac{1}{\varepsilon_0} \left(1 + (\gamma+1) \frac{\phi}{c^2}\right) \rho + \Or(c^{-3}).
\end{equation}

The $a$ component of the Maxwell equations \eqref{eq:Maxwell_simpl} is:
\begin{align}
	- \mu_0 \frac{1}{\sqrt{-g}} j_a &= g^{\mu\rho} \partial_\mu \partial_\rho \mathcal A_a - g^{0\rho} \partial_a \partial_\rho \mathcal A_0 \nonumber \\
	&= g^{\mu\rho} \partial_\mu \partial_\rho \mathcal A_a - \left(1 - 2\frac{\phi}{c^2} + \Or(c^{-4})\right) \varepsilon_0 \mu_0 \partial_a \partial_t \phi_\text{el.}
\end{align}
With $g^{\mu\rho} \partial_\mu \partial_\rho \mathcal A_a = \left(1 + 2\gamma \frac{\phi}{c^2}\right) \Delta \mathcal A_a - c^{-2} \partial_t^2 \mathcal A_a + \Or(c^{-4}) = \left(1 + 2\gamma \frac{\phi}{c^2}\right) (\Delta - c^{-2} \partial_t^2) \mathcal A_a + \Or(c^{-4}) $, this is equivalent to
\begin{equation}
	(\Delta - c^{-2} \partial_t^2) \mathcal A_a = -\mu_0 \left[ \left(1 + (\gamma-1) \frac{\phi}{c^2}\right) j_a - \left(1 - 2(\gamma+1) \frac{\phi}{c^2}\right) \varepsilon_0 \partial_a \partial_t \phi_\text{el.} \right] + \Or(c^{-4}).
\end{equation}
Rewriting $j_a = (1 - 2\gamma \frac{\phi}{c^2}) j^a + \Or(c^{-4})$ and introducing the internal vector potential\footnote{It is a transverse vector field because of the Coulomb gauge condition.} $\vect{\mathcal A} = \vect{\mathcal A}^\perp$ in $\mathcal A_\mu = (-\phi_\text{el.}/c, \vect{\mathcal A}^\perp)$, the wave equation for this potential reads as follows:
\begin{equation} \label{eq:wave_mag_pot}
	(\Delta - c^{-2} \partial_t^2) \vect{\mathcal A}^\perp = -\mu_0 \left[ \left(1 - (\gamma+1) \frac{\phi}{c^2}\right) \vect j - \left(1 - 2(\gamma+1) \frac{\phi}{c^2}\right) \varepsilon_0 \vect\nabla \partial_t \phi_\text{el.} \right] + \Or(c^{-4}).
\end{equation}

Now solving the equations \eqref{eq:Poisson_el_pot} and \eqref{eq:wave_mag_pot} for the internal potentials as in appendix A of \cite{sonnleitner18}, we see that to leading post-Newtonian order we arrive at the same potentials as in the non-gravitational case, up to prefactors: We have
\begin{equation} \label{eq:el_pot}
	\phi_\text{el.} = \left(1 + (\gamma+1) \frac{\phi}{c^2}\right) \phi_\text{el.,ng}
\end{equation}
and
\begin{equation} \label{eq:mag_pot}
	\vect{\mathcal A}^\perp = \left(1 - (\gamma+1) \frac{\phi}{c^2}\right) \vect{\mathcal A}_\text{ng}^\perp
\end{equation}
where the suffix `ng' stands for `non-gravitational', i.e. for the solutions from \eqref{eq:el_pot_orig}, \eqref{eq:mag_pot_orig}. Since we are assuming the absence of external charges, the external electric scalar potential $A_0$ vanishes due to the `external' equivalent of \eqref{eq:Maxwell_zero_comp}; and we introduce the external transverse vector potential $\vect A = \vect A^\perp$ as $A_\mu = (0, \vect A^\perp)$.

%%%%%%%%%%%%%%%%%%%%%%%%%%%%%%%%%%%%%%%%%%%%%%%%%%%%%%%%%%%
\subsection{Computation of the electromagnetic Lagrangian}
We will now use the obtained potentials to compute the electromagnetic Lagrangian
\begin{equation} \label{eq:L_em}
	L_\text{em} = \int\D^3\vect x \, \left(-\frac{1}{4\mu_0} \sqrt{-g} F_{\text{tot.}\mu\nu}F_\text{tot.}^{\mu\nu} + j^\mu A_{\text{tot.}\mu}\right),
\end{equation}
which follows from the action \eqref{eq:em_action}. For the internal kinetic Maxwell term, we obtain
\begin{align} \label{eq:kinetic_Maxwell_int}
	-\frac{1}{4\mu_0} \int\D^3\vect x \, \sqrt{-g} \mathcal F_{\mu\nu} \mathcal F^{\mu\nu} &= -\frac{1}{2\mu_0} \int\D^3\vect x \, \sqrt{-g} \partial_\mu \mathcal A_\nu \mathcal F^{\mu\nu} \nonumber \\
	\text{(P.I.)} \quad &= \frac{1}{2\mu_0} \int\D^3\vect x \, \sqrt{-g} \mathcal A_\nu \nabla_\mu \mathcal F^{\mu\nu} \nonumber
		\\&\quad- \frac{1}{2\mu_0} \int\D^3\vect x \, (\sqrt{-g} \partial_0 \mathcal A_\nu \mathcal F^{0\nu} + \mathcal A_\nu \partial_0 (\sqrt{-g} \mathcal F^{0\nu})) \nonumber \\
	&= -\frac{1}{2} \int\D^3\vect x \, \mathcal A_\nu j^\nu \nonumber
		\\&\quad- \frac{1}{2\mu_0} \int\D^3\vect x \, \Big(\sqrt{-g} \partial_0 \mathcal A_a (\nabla^0 \mathcal A^a - \nabla^a \mathcal A^0) \nonumber
		\\&\quad\qquad+ \mathcal A_a \partial_0 (\sqrt{-g} \nabla^0 \mathcal A^a - \sqrt{-g} \nabla^a \mathcal A^0)\Big) \nonumber \\
	\text{(P.I., $\partial_i g_{\mu\nu} = 0$, $\nabla^i \mathcal A_i = 0$)} \quad &= -\frac{1}{2} \int\D^3\vect x \, \mathcal A_\nu j^\nu \nonumber
		\\&\quad- \frac{1}{2\mu_0} \int\D^3\vect x \, \sqrt{-g} \left(\partial_0 \mathcal A_a \partial^0 \mathcal A^a + \mathcal A_a \partial_0\partial^0 \mathcal A^a\right) + \Or(c^{-4})
\end{align}
where in the last step, to perform the partial integration, we used that the metric is diagonal up to $\Or(c^{-4})$. Inserting the PPN metric, we obtain
\begin{align} \label{eq:kinetic_Maxwell_int_part_metric}
	&\hspace{-1.5em}-\frac{1}{2\mu_0} \int\D^3\vect x \, \sqrt{-g} \left(\partial_0 \mathcal A_a \partial^0 \mathcal A^a + \mathcal A_a \partial_0\partial^0 \mathcal A^a\right) \nonumber \\
	&= - \frac{\varepsilon_0}{2} \int\D^3\vect x \, \sqrt{-g} g^{00} g^{ab} \left(\partial_t \mathcal A_a \partial_t \mathcal A_b + \mathcal A_a \partial_t^2 \mathcal A_b\right) + \Or(c^{-4}) \nonumber \\
	&= \left(1 - (\gamma+1) \frac{\phi}{c^2}\right) \frac{\varepsilon_0}{4} \int\D^3\vect x \, \partial_t^2 \vect{\mathcal A}^2 + \Or(c^{-4}).
\end{align}
Thus, combining \eqref{eq:kinetic_Maxwell_int} and \eqref{eq:kinetic_Maxwell_int_part_metric}, the `purely internal' contribution of electromagnetism to the Lagrangian, including the explicit coupling term of the internal potential to the current, is:
\begin{align} \label{eq:L_em_int}
	L_\text{em,int.} &= \int\D^3\vect x \, \left( -\frac{1}{4\mu_0} \sqrt{-g} \mathcal F_{\mu\nu} \mathcal F^{\mu\nu} + j^\mu \mathcal A_\mu \right) \nonumber \\
	&= \frac{1}{2} \int\D^3\vect x \, j^\mu \mathcal A_\mu + \left(1 - (\gamma+1) \frac{\phi}{c^2}\right) \frac{\varepsilon_0}{4} \int\D^3\vect x \, \partial_t^2 \vect{\mathcal A}^2 + \Or(c^{-4})
\end{align}
Following \cite[appendix B]{sonnleitner18}, we will neglect the second integral in this expression\footnote
	{Quoting \cite[appendix B]{sonnleitner18}, it leads to terms `proportional to the electrostatic energy of the atom divided by $m_i c^2$ times $|\vect r_j|^2/c^2$, which goes beyond our level of approximation'.}.

For computing the other (purely external and mixed external-internal) contributions to the total electromagnetic Lagrangian \eqref{eq:L_em}, it is easiest to first rewrite the total kinetic Maxwell term. Using the PPN metric, but not the gauge condition, and writing $A_{\text{tot.}\mu} = (-\phi_{\text{el.,tot.}}/c, \vect A_\text{tot.})$, we have:
\begin{align} \label{eq:kin_Maxwell_total}
	&\hspace{-1.5em} -\frac{1}{4 \mu_0} \sqrt{-g} F_{\text{tot.}\mu\nu} F_\text{tot.}^{\mu\nu} \nonumber \\
	&= \frac{\varepsilon_0}{2} \sqrt{-g} \bigg[-g^{00} g^{ab} (\partial_t A_{\text{tot.}a} + \partial_a \phi_{\text{el.,tot.}}) (\partial_t A_{\text{tot.}b} + \partial_b \phi_{\text{el.,tot.}}) \nonumber
		\\&\qquad- c^2 (g^{ab} g^{cd} - g^{ad} g^{cb}) \partial_a A_{\text{tot.}c} \partial_b A_{\text{tot.}d})\bigg] + \Or(c^{-4}) \nonumber \displaybreak[0] \\
	&= \frac{\varepsilon_0}{2} \bigg[\left(1 - (\gamma+1) \frac{\phi}{c^2}\right) (\partial_t \vect A_\text{tot.} + \vect\nabla \phi_{\text{el.,tot.}})^2 \nonumber
		\\&\qquad- c^2 \left(1 + (\gamma+1) \frac{\phi}{c^2}\right) (\vect\nabla \times \vect A_\text{tot.})^2 \bigg] + \Or(c^{-4})
\end{align}
We recall that, as explained in section \ref{sec:Situation-CS-Conventions}, the boldface nabla symbol $\vect\nabla$ denotes the three-tuple of \emph{partial} derivatives, and $\vect\nabla \times \vect A_\text{tot.}$ denotes the `component-wise curl' of $\vect A_\text{tot.}$. Both of these operations are well-defined (independent of coordinates) once we have introduced the background structures (flat Minkowski metric and preferred time-like vector field).

The internal-internal term of \eqref{eq:kin_Maxwell_total} was considered above in \eqref{eq:L_em_int}. The external-external term gives ($\phi_\text{el.,tot.}$ is internal, and $\vect A = \vect A^\perp$ is transverse because of the gauge condition)
\begin{align} \label{eq:L_em_ext}
	L_\text{em,ext.} &= -\frac{1}{4\mu_0} \int\D^3\vect x \, \sqrt{-g} F_{\mu\nu} F^{\mu\nu} \nonumber \\
	&= \frac{\varepsilon_0}{2} \int\D^3\vect x \, \bigg[ \left(1 - (\gamma+1) \frac{\phi}{c^2}\right) (\partial_t \vect A^\perp)^2 \nonumber
		\\&\qquad- c^2 \left(1 + (\gamma+1) \frac{\phi}{c^2}\right) (\vect\nabla \times \vect A^\perp)^2 \bigg] + \Or(c^{-4}),
\end{align}
and the external-internal mixed term plus the interaction of the external potential with the current is (using partial integration and the gauge condition to get rid of the $\partial_t \vect A^\perp \cdot \vect\nabla \phi_\text{el.}$ term)
\begin{align} \label{eq:L_em_ext_int}
	L_\text{em,ext.-int.} &= \int\D^3\vect x \, j^\mu A_\mu - \frac{1}{2\mu_0} \int\D^3\vect x \, \sqrt{-g} \mathcal F_{\mu\nu} F^{\mu\nu} \nonumber \\
	&= \int\D^3\vect x \, j^\mu A_\mu  + \varepsilon_0 \int\D^3\vect x \, \bigg[ \left(1 - (\gamma+1) \frac{\phi}{c^2}\right) (\partial_t \vect{\mathcal A}^\perp) \cdot (\partial_t \vect A^\perp) \nonumber
		\\&\qquad- c^2 \left(1 + (\gamma+1) \frac{\phi}{c^2}\right) (\vect\nabla \times \vect{\mathcal A}^\perp) \cdot (\vect\nabla \times \vect A^\perp) \bigg] + \Or(c^{-4}).
\end{align}
Following \cite[appendix B]{sonnleitner18}, we will neglect the second integral also in this expression, since it is related to formally diverging backreaction terms.

Adding the Lagrangians \eqref{eq:L_em_int}, \eqref{eq:L_em_ext} and \eqref{eq:L_em_ext_int}, the total post-Newtonian electromagnetic Lagrangian (with the above-mentioned neglections following \cite{sonnleitner18}) reads:
\begin{align} \label{eq:Lagrangian_em_grav}
	L_\text{em} &= \frac{1}{2} \int\D^3\vect x \, (\vect j \cdot \vect{\mathcal A}^\perp - \rho \phi_\text{el.}) + \int\D^3\vect x \, \vect j \cdot \vect A^\perp \nonumber
	\\&\quad+ \frac{\varepsilon_0}{2} \int\D^3\vect x \, \left[ \left(1 - (\gamma+1) \frac{\phi}{c^2}\right) (\partial_t \vect A^\perp)^2 - c^2 \left(1 + (\gamma+1) \frac{\phi}{c^2}\right) (\vect\nabla \times \vect A^\perp)^2 \right] + \Or(c^{-4})
\end{align}
Inserting the internal potentials \eqref{eq:el_pot}, \eqref{eq:mag_pot} as well as the current and charge densities \eqref{eq:current_dens}, \eqref{eq:charge_dens} and dropping infinite self-interaction terms for the internal part of the electromagnetic Lagrangian we obtain
\begin{align} \label{eq:Lagrangian_em_grav_internal_explicit}
	&\hspace{-1.5em}\frac{1}{2} \int\D^3\vect x \, (\vect j \cdot \vect{\mathcal A}^\perp - \rho \phi_\text{el.}) \nonumber \\
	&= -\left(1 + (\gamma+1) \frac{\phi}{c^2}\right) \frac{e_1 e_2}{4\pi \varepsilon_0 r} \nonumber
		\\&\quad+ \left(1 - (\gamma+1) \frac{\phi}{c^2}\right) \frac{e_1 e_2}{8\pi \varepsilon_0 c^2} \left[ \frac{\dot{\vect r}_1 \cdot \dot{\vect r}_2}{r} + \frac{(\dot{\vect r}_1 \cdot \vect r) (\dot{\vect r}_2 \cdot \vect r)}{r^3} \right] + \Or(c^{-4}) \nonumber \\
	&= -\left(1 + (\gamma+1) \frac{\phi}{c^2}\right) \frac{e_1 e_2}{4\pi \varepsilon_0 r} + \frac{e_1 e_2}{8\pi \varepsilon_0 c^2} \left[ \frac{\dot{\vect r}_1 \cdot \dot{\vect r}_2}{r} + \frac{(\dot{\vect r}_1 \cdot \vect r) (\dot{\vect r}_2 \cdot \vect r)}{r^3} \right] + \Or(c^{-4}).
\end{align}

%%%%%%%%%%%%%%%%%%%%%%%%%%%%%%%%%%%%%%%%%%%%%%%%%%%%%%%%%%%
\section{The total Hamiltonian including all interactions}
\label{sec:Hamiltonian}
In this section we collect all previous findings
and combine them into the total Hamiltonian that 
characterises the dynamics of our two-particle system 
that is now also exposed to a non-trivial gravitational 
field. We will see that the Hamiltonian suffers various 
`corrections' as compared to the gravity-free case, and 
that these terms acquire an intuitive interpretation if
re-expressed in terms of the physical spacetime metric 
$g$.

%%%%%%%%%%%%%%%%%%%%%%%%%%%%%%%%%%%%%%%%%%%%%%%%%%%%%%%%%%%
\subsection{Computation of the Hamiltonian} 
\label{sec:Hamiltonian-Comp}
We will now compute the total Hamiltonian describing the 
atom in external electromagnetic and gravitational fields 
by repeating the calculation from section 
\ref{sec:Coupling-GF-Particle} while including the `gravitational 
corrections' to electromagnetism obtained in section~\ref{sec:Coupling-GF-EMF}.

Comparing the gravitationally corrected electromagnetic Lagrangian \eqref{eq:Lagrangian_em_grav}, \eqref{eq:Lagrangian_em_grav_internal_explicit} to the one without gravitational field ($\phi = 0$), we see that at our order of approximation the only differences occur in the external electromagnetic term and the internal Coulomb interaction term. Thus, when calculating the Hamiltonian, we have to only take care of these changes compared to the discussion of section \ref{sec:Coupling-GF-Particle}.

The canonical momentum conjugate to $\vect A^\perp$ is
\begin{equation}
	\vect\Pi^\perp = \frac{\delta L_\text{em}}{\delta(\partial_t \vect A^\perp)} = \varepsilon_0 \left(1 - (\gamma+1) \frac{\phi}{c^2}\right) \partial_t \vect A^\perp + \Or(c^{-4}),
\end{equation}
and the Hamiltonian for the external electromagnetic field thus is
\begin{align}
	H_\text{em,ext.} &= \int\D^3\vect x \, \vect\Pi^\perp \cdot \partial_t \vect A^\perp - L_\text{em,ext} \nonumber \\
	&= \frac{\varepsilon_0}{2} \int\D^3\vect x \, \left(1 + (\gamma+1) \frac{\phi}{c^2}\right) \left[ (\vect\Pi^\perp/\varepsilon_0)^2 + c^2 (\vect\nabla \times \vect A^\perp)^2 \right] + \Or(c^{-4}).
\end{align}

Thus, when including the new corrections, the classical Hamiltonian will be the same as given by \eqref{eq:Hamiltonian_class_no_em} and \eqref{eq:Hamiltonian_class_orig}, except for the Coulomb term and the external electromagnetic field energy both gaining the prefactor $\left(1 + (\gamma+1) \frac{\phi}{c^2}\right)$. The same will be true after quantising and performing the PZW transformation and electric dipole approximation: Using the same expression \eqref{eq:PZW} for the PZW operator, we arrive at the multipolar Hamiltonian from \eqref{eq:Hamiltonian_mult_no_em} and \eqref{eq:Hamiltonian_mult_orig}, except for this prefactor in the corresponding terms.

Now transforming to central and relative coordinates, as in going from \eqref{eq:Hamiltonian_mult_no_em} and \eqref{eq:Hamiltonian_mult_orig} to \eqref{eq:Hamiltonian_com_no_em} and \eqref{eq:Hamiltonian_com_orig}, the same stays true; we just have to keep in mind that the relation between the electromagnetic canonical momentum $\vect\Pi^\perp$ and the transverse external `coordinate electric field' $\vect E_\text{coord.}^\perp = -\partial_t\vect A^\perp$ now also gains an additional factor. Including this, the electric field--dipole interaction term again has the form $-\vect d \cdot \vect E_\text{coord.}^\perp(\vect R)$ as in the non-gravitational case, but the dipole polarisation self-interaction term $\frac{1}{2\varepsilon_0} \int\D^3\vect x \, {\vect{\mathcal P}_d^\perp}^2$ in $H_\text{AL}$ gains a prefactor (since it arises from the PZW-transformed external field energy).

Thus, comparing to the discussion of section \ref{sec:Coupling-GF-Particle}, the external electromagnetic field energy, the ${\vect{\mathcal P}_d^\perp}^2$ term in $H_\text{AL}$ and the Coulomb term are multiplied by the factor $\left(1 + (\gamma+1) \frac{\phi}{c^2}\right)$, and we arrive at the following total Hamiltonian:
\begin{subequations} \label{eq:Hamiltonian_com_final}
\begin{align}
	H_\text{[com],final} &= H_\text{C,final} + H_\text{A,final} + H_\text{AL,final} + H_\text{L,final} + H_\text{X}\\
	H_\text{C,final} &= H_\text{C,new} - \frac{1}{c^2} \frac{e^2}{4\pi \varepsilon_0 r} \phi(\vect R) \label{eq:Hamiltonian_com_C_final} \\
	H_\text{A,final} &= H_\text{A,new} - \gamma \frac{1}{c^2} \frac{e^2}{4\pi \varepsilon_0 r} \phi(\vect R) \label{eq:Hamiltonian_com_A_final} \\
	H_\text{AL,final} &= H_\text{AL} + \frac{1}{2\varepsilon_0} \int\D^3\vect x \, (\gamma+1) \frac{\phi}{c^2} {\vect{\mathcal P}_d^\perp}^2 \label{eq:Hamiltonian_com_AL_final} \\
	H_\text{L,final} &= \frac{\varepsilon_0}{2} \int\D^3\vect x \, \left(1 + (\gamma+1) \frac{\phi}{c^2}\right) \left[ (\vect\Pi^\perp/\varepsilon_0)^2 + c^2 (\vect\nabla \times \vect A^\perp)^2 \right] \label{eq:Hamiltonian_com_L_final}
\end{align}
\end{subequations}
Here, we have included the term $-\gamma \frac{1}{c^2} \frac{e^2}{4\pi \varepsilon_0 r} \phi(\vect R)$ into $H_\text{A,final}$ since it can be combined with the original Coulomb term from $H_\text{A}$ into
\begin{equation} \label{eq:Coulomb_metr}
	-\frac{e^2}{4\pi \varepsilon_0 r} \left(1 + \gamma\frac{\phi(\vect R)}{c^2}\right) = -\frac{e^2}{4\pi \varepsilon_0 \sqrt{{^{(3)}g_{\vect R}} (\vect r, \vect r)}} \; ,
\end{equation}
i.e. a Coulomb term expressed with the correct, metric relative distance.

Note that we neglected the terms $\vect r \cdot \vect\nabla\phi(\vect R)$ (since we assumed $\phi$ to be constant over the extension of the atom), thus arriving at the same cross terms $H_\text{X}$ as in \eqref{eq:Hamiltonian_com_cross_orig}; similarly, the corresponding term from $H_\text{A,new}$ \eqref{eq:Hamiltonian_com_atom_no_em} can be neglected.

To correctly interpret the atom--light interaction Hamiltonian \eqref{eq:Hamiltonian_com_AL_final}, one has to keep in mind that the field variables $\vect E^\perp$ and $\vect B$ in $H_\text{AL}$ \eqref{eq:Hamiltonian_com_AL_orig} are the \emph{coordinate} components $-\partial_t\vect A^\perp$ and $\vect\nabla \times \vect A^\perp$ respectively, which do not refer to an orthonormal frame in the physical spacetime metric $g$ in the presence of gravitational fields. This issue will be discussed in more detail in section \ref{sec:Hamiltonian-EM-ON}.

Since the cross terms $H_\text{X}$ are the same as in \cite{sonnleitner18}, we could now introduce new coordinates $\vect Q,\vect q,\vect p$ literally as in \eqref{eq:coords_decoup_orig} to eliminate these cross terms. Since the gravitational correction terms are of order $\Or(c^{-2})$, for them this change into the new coordinates would just amount to the replacements $\vect R \to \vect Q, \vect r \to \vect q, \vect p_{\vect r} \to \vect p$ at our order of approximation. Since it will not alter the following discussion, we will not perform this coordinate change in order to avoid adding an extra layer of potentially confusing notation.

%%%%%%%%%%%%%%%%%%%%%%%%%%%%%%%%%%%%%%%%%%%%%%%%%%%%%%%%%%%
\subsection{The system as a composite point particle}
\label{sec:Hamiltonian-CompositeParticle}
Fully writing out the central and internal Hamiltonian \eqref{eq:Hamiltonian_com_C_final}, \eqref{eq:Hamiltonian_com_A_final}, we obtain
\begin{align} \label{eq:Hamiltonian_com_C_final_full}
	H_\text{C,final} &= \frac{\vect P^2}{2M} \left[1 - \frac{1}{Mc^2} \left(\frac{\vect p_{\vect r}^2}{2\mu} + \frac{e_1 e_2}{4\pi \varepsilon_0 r}\right)\right] + \left[M + \frac{1}{c^2} \left(\frac{\vect p_{\vect r}^2}{2\mu} + \frac{e_1 e_2}{4\pi \varepsilon_0 r}\right) \right] \phi(\vect R) \nonumber
		\\&\quad - \frac{\vect P^4}{8M^3 c^2} + \frac{2\gamma+1}{2Mc^2} \vect P \cdot \phi(\vect R) \vect P + (2\beta-1) \frac{M \phi(\vect R)^2}{2 c^2} \; ,\\
	\label{eq:Hamiltonian_com_A_final_full}
	H_\text{A,final} &= \frac{{^{(3)}g^{-1}_{\vect R}} (\vect p_{\vect r}, \vect p_{\vect r})}{2\mu} + \frac{e_1 e_2}{4\pi \varepsilon_0 \sqrt{{^{(3)}g_{\vect R}} (\vect r, \vect r)}} \nonumber
		\\&\quad- \frac{m_1^3 + m_2^3}{M^3} \frac{\vect p_{\vect r}^4}{8\mu^3 c^2} + \frac{e_1 e_2}{4\pi \varepsilon_0} \frac{1}{2\mu M c^2} \left(\vect p_{\vect r} \cdot \frac{1}{r} \vect p_{\vect r} + \vect p_{\vect r} \cdot \vect r \frac{1}{r^3} \vect r \cdot \vect p_{\vect r}\right) \; ,
\end{align}
where we combined the gravitational correction terms in $H_\text{A,final}$ into metrically defined kinetic energy and Coulomb terms as in \eqref{eq:internal_kin_metr}, \eqref{eq:Coulomb_metr}.

Comparing to the Hamiltonian of a single point particle of mass $m$ in the PPN metric,
\begin{equation}
	H_\text{point}(\vect P, \vect R; m) = \frac{\vect P^2}{2m} + m \phi(\vect R) - \frac{\vect P^4}{8m^3 c^2} + \frac{2\gamma+1}{2mc^2} \vect P \cdot \phi(\vect R) \vect P + (2\beta-1) \frac{m \phi(\vect R)^2}{2 c^2} \; ,
\end{equation}
we thus see that the central Hamiltonian has, up to (and including) $\Or(c^{-2})$, exactly this form, with the mass $m$ replaced by $M + \frac{H_\text{A,final}}{c^2}$,
\begin{equation} \label{eq:Hamiltonian_composite_point}
	H_\text{C,final} = H_\text{point}\left(\vect P, \vect R; M + \frac{H_\text{A,final}}{c^2}\right),
\end{equation}
as could be naively expected from mass--energy equivalence. Thus, starting from first principles, we have shown that \emph{the system behaves as a `composite point particle' whose (inertial as well as gravitational) mass is comprised of the rest masses of the constituent particles as well as the internal energy.}

Note that this conclusion depends on the identification 
of terms as being `kinetic' and `interaction' energies, which 
in turn depends on the metric structure in their 
expressions. Had we not rewritten the internal kinetic 
energy \eqref{eq:internal_kin_metr} and the Coulomb interaction
\eqref{eq:Coulomb_metr} in terms of the physical metric 
$g$, the above conclusion could not have resulted.
Rather, we would have had to replace the inertial 
mass of the `composite particle' by 
$M + \frac{H_\text{A}}{c^2}$ and the gravitational mass 
by $M + (2\gamma+1) \frac{\vect p^2}{2\mu c^2} 
+ (\gamma+1) \frac{e_1 e_2}{4\pi \varepsilon_0 q c^2} 
= M + \frac{H_\text{A}}{c^2} + \gamma 
\left(2 \frac{\vect p^2}{2\mu c^2} 
+ \frac{e_1 e_2}{4\pi \varepsilon_0 q c^2}\right)$, 
which one could have erroneously interpreted as 
a violation of some naive form of the weak 
Equivalence Principle. But, clearly, such a 
conclusion would be premature, for it is based on 
the identification of terms -- like inertial and 
gravitational mass -- that is itself ambiguous.  
That ambiguity is here seen as a dependence on the 
background structure, which is used to define 
distances of positions and squares of momenta. Once 
these quantities are measured with the physical 
metric $g$, ambiguities and apparent conflicts 
with naive expectations disappear. That point 
has also been made in \cite{zych19}.

The quantities $\vec p'^2$ and $r'$ entering the 
Hamiltonian in \cite[eq. (18)]{zych19}, which are, 
in the language of \cite{zych19}, the square of the 
internal momentum and the distance `in the CM rest frame', 
are nothing but the geometric expressions 
${^{(3)}g^{-1}_{\vect R}} (\vect p_{\vect r}, \vect p_{\vect r})$ 
and $\sqrt{{^{(3)}g_{\vect R}} (\vect r, \vect r)}$ from above, 
measured using the physical metric of space. The internal Hamiltonian 
\eqref{eq:Hamiltonian_com_A_final_full} thus consists of 
kinetic and Coulomb interaction energies in terms of the 
physical geometry, in agreement with the expressions 
from \cite{zych19}, as well as the expected 
special-relativistic and `Darwin' corrections\footnote
	{Since these corrections are themselves of order $1/c^2$, the deviations of the physical from the flat metric do not enter here.}.

%%%%%%%%%%%%%%%%%%%%%%%%%%%%%%%%%%%%%%%%%%%%%%%%%%%%%%%%%%%
\subsection{The electromagnetic expressions in terms of components with respect to orthonormal frames}
\label{sec:Hamiltonian-EM-ON}
The expressions derived above in \eqref{eq:Hamiltonian_com_final} include components 
of the electromagnetic field with respect to 
coordinates which, albeit not chosen arbitrarily, 
have no direct metric significance. We recall that 
we used coordinates that are adapted to the chosen 
background structure $(\eta,u)$. This means that 
$u = \partial/\partial t$ and 
$\eta = \eta_{\mu\nu} \, \D x^\mu \otimes \D x^\nu$ 
with $(\eta_{\mu\nu}) = \mathrm{diag}(-1,1,1,1)$.
The corresponding local reference frames 
$\{\partial_\mu := \partial/\partial x^\mu \mid \mu
= 0,1,2,3\}$ are orthonormal with respect to $\eta$, 
but not with respect to the physical metric $g$.

In this subsection we will re-express our findings 
in terms of components with respect to orthonormal frames with 
respect to $g$, which we will call the `physical components', as opposed to the `coordinate 
components' used so far. We stress that, despite 
this terminology, there is nothing wrong or 
`unphysical' with representing fields in terms 
of components of non orthonormal bases, as long 
as the metric properties are spelled out at the 
same time. Yet it is clearly convenient to be 
able to read off metric properties, which 
bear direct physical significance, from the 
expressions involving the components alone, without 
at the same time having to recall the values of the 
metric components as well.

The full atom--light interaction Hamiltonian $H_\text{AL,final}$ \eqref{eq:Hamiltonian_com_AL_final} 
reads
\begin{align} \label{eq:Hamiltonian_com_AL_final_full}
	H_\text{AL,final} &= - \vect d \cdot \vect E_\text{coord.}^\perp(\vect R) + \frac{1}{2M} \{\vect P \cdot [\vect d \times \vect B_\text{coord.}(\vect R)] + \text{H.c.}\} \nonumber
		\\&\quad - \frac{m_1 - m_2}{4 m_1 m_2} \{\vect p_{\vect r} \cdot [\vect d \times \vect B_\text{coord.}(\vect R)] + \text{H.c.}\} \nonumber
		\\&\quad + \frac{1}{8\mu} (\vect d \times \vect B_\text{coord.}(\vect R))^2 + \frac{1}{2\varepsilon_0} \int\D^3\vect x \, \left(1 + (\gamma+1) \frac{\phi}{c^2}\right) {\vect{\mathcal P}_d^\perp}^2 (\vect x, t) \; ,
\end{align}
where $\vect E_\text{coord.}^\perp = - \partial_t \vect A^\perp$ and 
$\vect B_\text{coord.}^\perp = \vect\nabla \times \vect A^\perp$, 
i.e. $E_{\text{coord.}a} = c F_{a0}$ and 
$B_\text{coord.}^a = \varepsilon^{abc} F_{bc}$, are given by the 
coordinate components of the electromagnetic field tensor 
that refer to non orthonormal frames $\partial_\mu$. 
Since our metrics $\eta$ and $g$ are both diagonal in 
the coordinates used and to the order of approximation 
employed here, we simply need to divide 
$\partial_\mu$ by the square-root of the 
modulus of $g_{\mu\mu}$ (no summation) in order 
to get an orthonormal frame $\E_{\ul{\mu}}$ for $g$, called a 
`tetrad': 
\begin{subequations} \label{eq:tetrad}
\begin{align}
	\E_{\ul{0}} = \frac{1}{\sqrt{-g_{00}}} \partial_0 &= \left(1 - \frac{\phi}{c^2}\right) \partial_0 \; , \\
	\E_{\ul{a}} = \frac{1}{\sqrt{g_{aa}}} \partial_a &= \left(1 + \gamma \frac{\phi}{c^2}\right) \partial_a
\end{align}
\end{subequations}
Note that from now on underlined indices refer to 
the tetrad (`physical components'), non underlined 
ones to the coordinate basis (`coordinate components').

The `physical components' of the electric and magnetic 
fields with respect to the tetrad 
are given in terms of the tetrad components 
of the field tensor through   
\begin{equation} \label{eq:EM_fields_tetrad}
E_{\text{phys.}\ul{a}} 
 = c F_{\ul{a} \ul{0}} \, , \qquad
B_{\text{phys.}}^{\ul{a}} 
 = \sum_{b,c = 1}^3 \varepsilon^{abc} F_{\ul{b}\ul{c}} \, ,
\end{equation}
where $\varepsilon^{abc}$ is the totally antisymmetric symbol. 
This can be understood in a more geometric way: Denoting by 
$\tilde\varepsilon^{\ul{a}\ul{b}\ul{c}}$ the tetrad components 
of the spatial volume form $\tilde\varepsilon$ that is induced 
by the physical metric $g$ (instead of the spatial volume 
form $\varepsilon$ induced by the background metric $\eta$),
we have the numerical identity 
\begin{equation}
	\tilde\varepsilon^{\ul{a}\ul{b}\ul{c}} = \varepsilon^{abc}
\end{equation}
since the tetrad basis vectors $\{\E_{\ul{a}}\}$ are 
orthonormal with respect to $g$. Thus, the magnetic field 
components can be written as 
\begin{equation}
	B_{\text{phys.}}^{\ul{a}} = \tilde\varepsilon^{\ul{a}\ul{b}\ul{c}} F_{\ul{b}\ul{c}} \, ,
\end{equation}
with clear geometric meaning.
Inserting the tetrad \eqref{eq:tetrad}, the 
components \eqref{eq:EM_fields_tetrad} are 
related to the coordinate expressions 
$E_{\text{coord.}a} = c F_{a0}$ and 
$B_\text{coord.}^a = \varepsilon^{abc} F_{bc}$ by
\begin{equation}
	\vect E_\text{phys.} = \frac{1}{\sqrt{-g_{00}}} \left(1 + \gamma \frac{\phi}{c^2}\right) \vect E_\text{coord.} \; , \;
	\vect B_\text{phys.} = \left(1 + 2\gamma \frac{\phi}{c^2}\right) \vect B_\text{coord.} \;.	
\end{equation}

Similarly, the dipole moment $d^a = \sum_{k=1,2} e_k (r_k^a - R^a)$ is defined via coordinate distances, and thus the physically significant, metric dipole moment is given by
\begin{equation}
	d_\text{phys.}^{\ul{a}} = \E^{\ul{a}}_b d^b = \left(1 - \gamma \frac{\phi}{c^2}\right) d^a.
\end{equation}
Thus the electric dipole interaction term in the Hamiltonian takes the form
\begin{equation}
	- \vect d \cdot \vect E_\text{coord.}^\perp(\vect R) = - \sqrt{-g_{00}(\vect R)} \, \vect d_\text{phys.} \cdot \vect E_\text{phys.}^\perp(\vect R)
\end{equation}
when expressed in terms of physical components. The `gravitational time dilation' factor $\sqrt{-g_{00}}$ in this expression could now also be absorbed by referring the time evolution to the proper time of the observer situated at $\vect R$ instead of coordinate time \cite{marzlin95,laemmerzahl95}.

Similarly, all the other interaction terms from 
\eqref{eq:Hamiltonian_com_AL_final_full} can be 
rewritten in terms of tetrad components. The only 
difficulty arises when considering the R\"ontgen 
term, i.e. the second term in the interaction 
Hamiltonian, since it involves the momentum 
$\vect P$: If the components $P_a$ were just the 
components of a classical covector field (i.e. a 
one-form), there would be no problem in computing 
its tetrad components as
\begin{equation} \label{eq:momentum_phys}
	P_{\text{phys.}\ul{a}} = \E^b_{\ul{a}} P_b = \left(1 - \gamma \frac{\phi}{c^2}\right) P_a \; .
\end{equation}
However, the $P_a$ are operators that don't 
commute with the centre of mass position $\vect R$,
such that in the application of \eqref{eq:momentum_phys} 
one has to deal with with operator ordering ambiguities 
(which is, of course, a well-known issue regarding 
curvilinear coordinate transformations in Quantum Mechanics). 
Of course, to avoid dealing with these ambiguities, one can 
stay with the coordinate components of the momentum 
and rewrite only the \emph{other} quantities in 
terms of tetrad components, arriving at 
\begin{equation}
	\frac{1}{2M} \{\vect P \cdot [\vect d \times \vect B_\text{coord.}(\vect R)] + \text{H.c.}\} = \frac{1}{2M} \left\{\vect P \cdot \left(1 - \gamma \frac{\phi(\vect R)}{c^2}\right) [\vect d_\text{phys.} \times \vect B_\text{phys.}(\vect R)] + \text{H.c.}\right\}
\end{equation}
for the Röntgen term. When doing so, to give a well-defined 
geometric meaning to the resulting expression on the 
right-hand side, one has to keep in mind that the 
components $\vect P$ of the momentum refer to the 
coordinate basis and the components 
$\vect d_\text{phys.}, \vect B_\text{phys.}$ of the 
dipole moment and the magnetic field refer to the tetrad.

Note that for the internal momentum $\vect p_{\vect r}$, 
occurring in the third interaction term in \eqref{eq:Hamiltonian_com_AL_final_full}, 
such operator ordering ambiguities do not arise, since 
we assumed the gravitational potential $\phi$ to be constant 
over the extension of the atom.

To the best of our knowledge, the atom--light interaction terms in the 
presence of gravity obtained in \eqref{eq:Hamiltonian_com_AL_final_full} 
and discussed above are new, save for the electric dipole 
coupling which was already discussed in \cite{marzlin95,laemmerzahl95}.

Finally, expressing the external field energy 
\eqref{eq:Hamiltonian_com_L_final} in terms of 
tetrad components, we obtain
\begin{align}
	H_\text{L,final} &= \frac{\varepsilon_0}{2} \int\D^3\vect x \, \left(1 + (1 - 3\gamma) \frac{\phi}{c^2}\right) \left[ {\vect E_\text{phys.}^\perp}^2 + c^2 \vect B_\text{phys.}^2 \right] \nonumber \\
	&= \frac{\varepsilon_0}{2} \int\D^3\vect x \, \sqrt{-g} \left[ {\vect E_\text{phys.}^\perp}^2 + c^2 \vect B_\text{phys.}^2 \right]
\end{align}
which is the standard result of the flat-spacetime 
electromagnetic field energy \cite{jackson98} minimally coupled to 
gravity \cite{misner73}, as was to be expected.\footnote
	{This result would have been immediate if we did the whole calculation in terms of tetrad components instead of coordinate components, as would have some steps in the calculation of the electromagnetic Lagrangian. However, as stressed in section \ref{sec:Situation-CS-Conventions}, the approach based on the background structures enabled us to provide a direct comparison with the original calculation of \cite{sonnleitner18}.}

%%%%%%%%%%%%%%%%%%%%%%%%%%%%%%%%%%%%%%%%%%%%%%%%%%%%%%%%%%%
\section{A comment on the Equivalence Principle in Quantum Mechanics}
\label{sec:EP}
As advertised at the end of the introduction, we now 
wish to come back to the less technical but conceptually 
all-important issue surrounding the meaning of 
`Equivalence Principle', in particular concerning its 
susceptibility to be more accurately tested by genuine 
quantum matter. Clearly, that possibility presupposes
the formulation of the Equivalence Principle 
in a way that is fully compatible with 
Quantum Mechanics. Our discussion here is motivated 
by our impression that this need, simple and obvious 
as it might seem to be, is often unduly neglected. 
As emphasised before, our central point of concern 
results from our conviction that any generally valid 
implementation of the Equivalence Principle 
into Quantum Mechanics should not be based on 
\emph{a priori} assumptions concerning the state 
of the matter. In particular, it should not rely 
on semi-classical notions like `wordline', that often enter 
discussions in a fundamental way%
\footnote{In that respect we fully agree with 
\cite{anastopoulos18}.}; 
see, e.g., \cite{zych11,pikovski15,roura18,giese19,loriani19,zych19}. 
It should also not 
rely on other classical formulations, like 
`equality of inertial and gravitational mass', 
for reasons discussed below.

Rather, the meaning of the 
Equivalence Principle is to ensure that \emph{a single}
spacetime geometry accounts for \emph{all} matter 
couplings, thereby comprising all gravity phenomena on
all kinds of matter; compare \cite{thorne73}. 
This does clearly not imply that all `things' 
(bodies, wave packets, \ldots) `fall' with the same 
acceleration. First, in order to give a measure 
for the acceleration of `fall', we need to 
kinematically define that term; that is, we need to 
specify a structure on which we can read-off such 
quantities as acceleration. In other words: we need to
define a worldline. As discussed before, 
quantum-mechanical systems only do that in 
very special states.
Second, even in case a worldline is approximately 
definable, there is no reason why it must coincide 
with that of another system if both start out in 
the same centre of mass initial state. This is not
even true in classical physics. The universality of
free fall reduces to the equality of worldlines only 
in case of `test particles', idealised objects of 
partially opposing properties%
\footnote{A `test particle' must have negligible 
size as compared to the curvature radius
of the background geometry, so as to not couple 
to second and higher derivatives of the metric (i.e.
no curvature couplings). At the same time, it cannot 
be too small, so as to not create a significant 
gravitational self-energy and an appreciable 
self-gravitational field on top of that of 
background spacetime. Moreover, it may have no 
non-vanishing charges, spin, or higher moments 
amongst its mass multipoles. This should make it clear 
that the notion of `test particle' is contextual and 
approximative. One and the same object may be a good 
test particle in one situation and badly fail to be 
such in another.}. 
For realistic bodies, which are 
spinning and/or possess higher mass multipoles, the 
centres of mass will not define universal 
point-particle trajectories (i.e. geodesics). 
To find the equations of motion for structured bodies
in arbitrary gravitational fields is, in fact, one of 
the most difficult challenges in General Relativity
research~\cite{puetzfeld15}. However, as long as 
all deviations from the geodesic motion find their 
explanations in couplings to the spacetime geometry, 
no violation of the Equivalence Principle should be 
concluded. Likewise, extended wave packets in Quantum 
Mechanics will move in a way that depends on the higher
moments of their probability distribution~\cite{visser17},
but this does not contradict any Equivalence 
Principle requirement, contrary to the impression 
given in \cite{visser17}. In that context we wish to 
remind the reader of our discussion in the introduction.

Finally, we also wish to comment on attempts to 
translate the equality of inertial and gravitational 
(more precisely: passive gravitational) mass into 
Quantum Mechanics; see, e.g., \cite{zych18}. 
These attempts also bear the danger 
to prematurely conclude apparent violations, even though 
all the couplings are through the geometry only. 
In fact, it seems to us that those attempts suffer from a 
logical flaw that we now wish to briefly characterise. 

First the positive results: It is an elementary theorem 
in ordinary Quantum Mechanics that, given a spatially
homogeneous but arbitrarily time dependent force field, 
any solution to the Schr\"odinger equation in that 
field corresponds to a solution of the force-free 
Schr\"odinger equation written in the coordinates of 
a frame that freely falls according to the classical 
equations of motion in the given force field. This 
is, e.g., proven in \cite{giulini12}. If the 
force field is that of Newtonian gravity, such that 
the classical trajectory only depends on the ratio 
of the inertial to the gravitational mass, then the motion 
of any wave packet is that of a free Schr\"odinger 
wave in a \emph{universally} (i.e. the same for all 
wave functions and all particles with the same ratio
of inertial and gravitational mass) 
falling frame. In that sense there is no observational 
difference between a falling wave packet in a 
homogeneous gravitational field and a free wave packet 
without gravity viewed from a frame that rigidly moves 
with equal but opposite acceleration to that determined 
by the gravity field. This precisely corresponds to the 
Equivalence Principle for homogeneous fields in Newtonian 
gravity which we therefore see to extend to Quantum Mechanics. 

In Classical Mechanics, we can extend the  universality 
of free fall to non-homogeneous fields if 
we restrict attention to pointlike test particles. Their 
equation of motion then guarantees the universality 
of free fall, given that their inertial equals their 
gravitational mass. This shows that we need the 
particular equations of motion to provide the logical 
link between the universality of free fall on one side 
and the equality of the inertial and gravitational 
mass on the other. As such the latter equality does not tell 
us anything if this link is not provided. That, in our 
opinion, is the flaw in any premature identification
of the weak Equivalence Principle with the equality of 
inertial and gravitational mass (leaving alone the ambiguity 
to define the latter in non-Newtonian situations). 
In particular this applies to Quantum Mechanics, 
where neither the notion of a `test particle' exists, 
nor does there exist an obvious worldline the equation 
of which could provide the sought-for link. So, for 
general states, it remains logically unclear what a 
possible equality of somehow defined inertial and 
gravitational masses implies for the universality of free 
fall, though it might make contextual sense for special 
states. However, that state dependency 
means that it cannot be used as a fundamental principle.

%%%%%%%%%%%%%%%%%%%%%%%%%%%%%%%%%%%%%%%%%%%%%%%%%%%%%%%%%%%
\section{Conclusions}
\label{sec:Conclusion}
In this paper we extended the calculation of 
\cite{sonnleitner18} of a Hamiltonian describing 
an electromagnetically interacting 
two-particle system to non-flat spacetimes representing 
weak gravitational fields in the sense of being close 
to Minkowski spacetime. Starting from first principles we 
performed a post-Newtonian expansion in terms 
of the inverse velocity of light that led to 
leading-order corrections comprising special- 
and general-relativistic effects. The former were 
fully encoded in  \cite{sonnleitner18}, but the 
latter are new. Our first principles were the 
special-relativistic action of two equally but 
oppositely charged particles with their 
electromagnetic interaction up to, and including, 
terms of order $c^{-2}$, the minimal coupling scheme 
for gravity, and canonical quantisation applied 
to the degrees of freedom in the Hamiltonian 
formalism. As in \cite{sonnleitner18} we neglected 
all terms of third and higher order in $c^{-1}$, which 
physically means that we neglected radiation-reaction 
and also that we avoided obstructions on the applicability 
of the Hamiltonian formalism that result from the 
infamous `no-interaction theorem' \cite{currie63},
whose impact only starts at the 6th order in a 
$c^{-1}$ expansion \cite{martin78}.

As for \cite{sonnleitner18} in the gravity-free case, 
we see the virtue of our calculation in its firm rooting
in explicitly spelled out principles, that leave no 
doubt concerning the questions of consistency and 
completeness of relativistic corrections. This, in 
our opinion, distinguishes our work from previous ones 
by other authors, who were also concerned with the 
coupling of composite particle quantum systems -- like 
atoms or molecules -- to external gravitational fields,
who phrase their account of relativistic corrections 
in terms of semi-classical notions, like smooth 
worldlines and comparisons of their associated lengths 
(i.e. `proper time' and `redshift'); e.g. 
\cite{zych11,pikovski15,roura18,giese19,loriani19,zych19}. 
We emphasised in the introduction 
and also in our discussion of the Equivalence Principle 
that answers to the fundamental question of 
gravity--matter coupling in Quantum Mechanics should 
not be based on \emph{a priori} restricted states that 
imply a semi-classical behaviour of some of the 
(factorising) degrees of freedom. Rather, they should 
apply to all states in an equally valid fashion.

Similar to the gravity-free case, we now \emph{derived}
the result that the centre of mass motion of the system 
can be viewed as that of a `composite point particle', including 
in its mass the internal energy of the system. This result may 
be anticipated in a heuristic fashion on semi-classical 
grounds, but, as seen, its proper derivation requires 
some efforts. We stress once more 
that for this interpretation it was crucial to express the 
Hamiltonian in terms of the physical space-time metric. 
As a result, our work lends justification to current 
experimental proposals in atom interferometry that so far were based on these 
heuristic ideas on the basis of which completeness of 
the relativistic effects could not be reliably judged; 
e.g.~\cite{zych11,pikovski15,roura18,giese19,loriani19,zych19}. 
Moreover, it also applies to new experimental setups, 
like that of an ion trap in a gravitational field, 
currently under investigation~\cite{martinez_inprep} 
in extension of \cite{haustein19}. Finally we 
mention that due to our explicit parametrisation of the 
gravitational field by means of the Eddington--Robertson 
parameters, our formulae will also apply to possible 
quantum tests of General Relativity against test theories 
within that class.

%%%%%%%%%%%%%%%%%%%%%%%%%%%%%%%%%%%%%%%%%%%%%%%%%%%%%%%%%%%
\section*{Acknowledgements}
We thank Klemens Hammerer for pointing out 
reference \cite{sonnleitner18} and asking 
questions that initiated and encouraged the work 
reported here, as well as for critical reading of 
the manuscript and the suggestion of improvements.
This work was supported by the Deutsche
Forschungsgemeinschaft through the Collaborative 
Research Centre 1227 (DQ-mat), project B08. 
We thank all members for fruitful and encouraging 
discussions on topics related to the one dealt 
with in this paper.

%%%%%%%%%%%%%%%%%%%%%%%%%%%%%%%%%%%%%%%%%%%%%%%%%%%%%%%%%%%
\appendix

%%%%%%%%%%%%%%%%%%%%%%%%%%%%%%%%%%%%%%%%%%%%%%%%%%%%%%%%%%%
\section{Formulae from the paper of Sonnleitner and Barnett} \label{app:sonnleitner_formulae}

\newcommand{\SBtag}[1]{\tag{\cite{sonnleitner18}.#1}}
\newcommand{\SBtagc}[1]{\tag{\cite{sonnleitner18}.#1$\star$}}
\newcommand{\corr}{($\star$)}
\newcommand{\red}[1]{\textcolor{red}{#1}}
\newcommand{\redbin}[1]{\mathbin{\textcolor{red}{#1}}}

In this appendix we reproduce all formulae from \cite{sonnleitner18} that are used in the main text, while providing a little context. We use the original numbering, prepended with `\cite{sonnleitner18}.'. For formulae containing an error in \cite{sonnleitner18}, we give here a corrected version; the corrections are highlighted in \red{red} and the number is marked with a star.

The classical Lagrangian for two particles interacting with electromagnetic potentials\footnote{In the absence of gravity, as this is the situation considered in \cite{sonnleitner18}.} is
\begin{align} \label{eq:Lagrangian_class_start_orig}
	L = &- \sum_{i=1,2} m_i c^2 \sqrt{1 - \dot{\vect r}_i^2 / c^2} + \int \D^3\vect x \, (\vect j \cdot \vect A_\text{tot} - \rho \phi_\text{tot}) \nonumber \\
	&+ \frac{\varepsilon_0}{2} \int \D^3\vect x \, [(\partial_t \vect A_\text{tot} + \vect \nabla \phi_\text{tot})^2 - c^2 (\vect\nabla \times \vect A_\text{tot})^2]. \SBtag{4}
\end{align}
Note that here, in the notation of \cite{sonnleitner18}, $\phi_\text{tot}$ is the total electric potential, not to be confused with our gravitational potential $\phi$ from the main text.

One then splits the electromagnetic potentials into internal (generated by the particles) and external parts, employs the Coulomb gauge and solves the Maxwell equations for the internal part in lowest order (see \eqref{eq:el_pot_orig}, \eqref{eq:mag_pot_orig}). Inserting the internal potential solutions and expanding the kinetic terms for the particles, one arrives at the post-Newtonian Lagrangian
\begin{align}
	& L(\vect r_1, \dot{\vect r}_1, \vect r_2, \dot{\vect r}_2, \vect A^\perp, \dot{\vect A}^\perp) \nonumber \\
	& \quad = L_\text{Darwin}(\vect r_1, \dot{\vect r}_1, \vect r_2, \dot{\vect r}_2) + \frac{\varepsilon_0}{2} \int \D^3\vect x \, [(\partial_t \vect A^\perp)^2 \nonumber \\
		& \qquad - c^2 (\vect\nabla \times \vect A^\perp)^2] + \int \D^3\vect x \, \vect j \cdot \vect A^\perp \; , \SBtag{8} \label{eq:Lagrangian_class_postNewt_orig} \\
	& L_\text{Darwin}(\vect r_1, \dot{\vect r}_1, \vect r_2, \dot{\vect r}_2) \nonumber \\
	& \quad = \frac{m_1 \dot{\vect r}_1^2}{2} + \frac{m_1 \dot{\vect r}_1^4}{8c^2} + \frac{m_2 \dot{\vect r}_2^2}{2} + \frac{m_2 \dot{\vect r}_2^4}{8c^2} \nonumber \\
		& \qquad - \frac{1}{4\pi\varepsilon_0} \frac{e_1 e_2}{r} \left(1 - \frac{\dot{\vect r}_1 \cdot \dot{\vect r}_2}{2c^2}\right) + \frac{e_1 e_2}{4\pi\varepsilon_0} \frac{(\dot{\vect r}_1 \cdot \vect r) (\dot{\vect r}_2 \cdot \vect r)}{2 r^3 c^2} \; , \SBtag{9} \label{eq:Lagrangian_class_Darwin_orig}
\end{align}
where $\vect r = \vect r_1 - \vect r_2$ and $r = |\vect r|$.

The classical Hamiltonian obtained by Legendre transforming this Lagrangian is
\begin{align} \label{eq:Hamiltonian_class_orig}
	H &= \frac{\bar{\vect p}_1^2}{2 m_1} \redbin{-} \frac{\bar{\vect p}_1^4}{8 m_1^3 c^2} + \frac{\bar{\vect p}_2^2}{2 m_2} \redbin{-} \frac{\bar{\vect p}_2^4}{8 m_2^3 c^2} + \frac{1}{4\pi\varepsilon_0} \frac{e_1 e_2}{r} \left(1 - \frac{\bar{\vect p}_1 \cdot \bar{\vect p}_2}{2 m_1 m_2 c^2}\right) \nonumber
		\\&\quad - \frac{e_1 e_2}{4\pi\varepsilon_0} \frac{(\bar{\vect p}_1 \cdot \vect r)(\bar{\vect p}_2 \cdot \vect r)}{2 r^3 c^2 \red{m_1 m_2}} + \frac{\varepsilon_0}{2} \int \D^3\vect x \, [(\vect\Pi^\perp / \varepsilon_0)^2 + c^2(\vect\nabla \times \vect A^\perp)^2], \SBtagc{12}
\end{align}
where $\bar{\vect p}_i = \vect p_i \redbin{-} e_i \vect A^\perp(\vect r_i)$ \corr.

The PZW transformation operator is
\begin{equation} \SBtagc{14} \label{eq:PZW}
	U = \mathrm{e}^{-\I\Lambda} = \exp\left[\redbin{-} \frac{\I}{\hbar} \int\D^3\vect x \, \vect{\mathcal P}(\vect x, t) \cdot \vect A^\perp(\vect x, t)\right],
\end{equation}
where $\vect{\mathcal P}$ is the polarisation field
\begin{align}
	\vect{\mathcal P}(\vect x, t) &= \sum_{i=1,2} e_i [\vect r_i(t) - \vect R(t)] \nonumber \\
		&\qquad \times \int_0^1 \D\lambda \, \delta\{\vect x - \vect R(t) - \lambda [\vect r_i(t) - \vect R(t)]\}. \SBtag{15}
\end{align}
The transformation amounts to the following change of canonical momenta:
\begin{align}
	\vect p_i &\to U \vect p_i U^\dagger = \vect p_i + \hbar \vect\nabla_{\vect r_i} \Lambda, \SBtag{19a} \\
	\vect\Pi^\perp(\vect x) &\to \vect\Pi^\perp(\vect x) \redbin{+} \vect{\mathcal P}^\perp(\vect x). \SBtagc{19b}
\end{align}
In electric dipole approximation, i.e. expanding to first order in $\bar{\vect r}_i := \vect r_i - \vect R$, and using $\sum_{j=1,2} e_j = 0$, we find
\begin{align*} \label{eq:SB_21}
	\hbar \vect\nabla_{\vect r_{1,2}} \Lambda &\simeq \red{e_{1,2}} [\vect A^\perp(\vect R) + (\bar{\vect r}_{1,2} \cdot \vect\nabla) \vect A^\perp(\vect R)]
		\\&\quad + \frac{e_1 \vect r_1 + e_2 \vect r_2}{2} \times [\vect\nabla \times \vect A^\perp(\vect R)]. \SBtagc{21}
\end{align*}
Thus, under the PZW transformation and the dipole approximation the momenta transform as $\vect p_i \redbin{-} e_i \vect A(\vect r_i) \to \vect p_i + \vect d \times \vect B(\vect R) / 2$ \corr, where $\vect d$ is the dipole moment.

Terms of the form
\begin{equation} \SBtag{22} \label{eq:SB_negl_interaction}
	\frac{\vect p_i \cdot [\vect d \times \vect B(\vect R)]}{m_i m_j c^2} \propto \frac{|\vect p_i|}{m_i c} \frac{|\vect d \cdot \vect E(\vect R)|}{m_j c^2}
\end{equation}
are neglected, since the atom--light interaction energy is assumed much smaller than the internal atomic energy, which is in turn much smaller than the rest energies of the particles. The multipolar Hamiltonian in electric dipole approximation is then
\begin{align} \label{eq:Hamiltonian_mult_orig}
	H_\text{[mult]} &\simeq \frac{[\vect p_1 + \frac{1}{2} \vect d \times \vect B(\vect R)]^2}{2 m_1} + \frac{[\vect p_2 + \frac{1}{2} \vect d \times \vect B(\vect R)]^2}{2 m_2} \nonumber
		\\&\quad - \frac{e^2}{4\pi\varepsilon_0 r} + \frac{\varepsilon_0}{2} \int \D^3\vect x \, [(\vect\Pi^\perp \redbin{+} \vect{\mathcal P}_d^\perp)^2 / \varepsilon_0^2 + c^2 \vect B^2] \nonumber
		\\&\quad \redbin{-} \frac{\vect p_1^4}{8 m_1^3 c^2} \redbin{-} \frac{\vect p_2^4}{8 m_2^3 c^2} + \frac{e^2}{16\pi\varepsilon_0 c^2 m_1 m_2} \nonumber
		\\&\quad \times \left[ \vect p_1 \cdot \frac{1}{r} \vect p_2 + (\vect p_1 \cdot \vect r) \frac{1}{r^3} (\vect r \cdot \vect p_2) + (1 \leftrightarrow 2) \right], \SBtagc{23}
\end{align}
where $\vect{\mathcal P}_d = \redbin{+} \vect d \delta(\vect x - \vect R)$ \corr\ is the polarisation in electric dipole approximation.

Expressed in Newtonian centre of mass coordinates, the Hamiltonian is as follows:
\begin{subequations}
	\makeatletter
	\def\@currentlabel{\cite{sonnleitner18}.25$\star$}
	\makeatother
	\label{eq:Hamiltonian_com_orig}
\begin{align*}
	H_\text{[com]} &= H_\text{C} + H_\text{A} + H_\text{AL} + H_\text{L} + H_\text{X}, \SBtag{25a} \\
	H_\text{C} &= \frac{\vect P^2}{2M} \left[1 - \frac{\vect P^2}{4M^2 c^2} - \frac{1}{M c^2} \left(\frac{\vect p_{\vect r}^{\red{2}}}{2\mu} - \frac{e^2}{4\pi\varepsilon_0 r}\right)\right], \SBtagc{25b} \\
	H_\text{A} &= \frac{\vect p_{\vect r}^2}{2\mu} \left(1 - \frac{m_1^3 + m_2^3}{M^3} \frac{\vect p_{\vect r}^2}{4 \mu^2 c^2}\right) - \frac{e^2}{4\pi\varepsilon_0}
		\\&\quad \times \left[\frac{1}{r} + \frac{1}{2\mu M c^2} \left( \vect p_{\vect r} \cdot \frac{1}{r} \vect p_{\vect r} + \vect p_{\vect r} \cdot \vect r \frac{1}{r^3} \vect r \cdot \vect p_{\vect r} \right)\right], \SBtag{25c} \\
	H_\text{AL} &= - \vect d \cdot \vect E^\perp(\vect R) + \frac{1}{2M} \{\vect P \cdot [\vect d \times \vect B(\vect R)] + \text{H.c.}\}
		\\&\quad - \frac{m_1 - m_2}{\red{4} m_1 m_2} \{\vect p_{\vect r} \cdot [\vect d \times \vect B(\vect R)] + \text{H.c.}\}
		\\&\quad + \frac{1}{8\mu} (\vect d \times \vect B(\vect R))^2 + \frac{1}{2\varepsilon_0} \int\D^3\vect x \, {\vect{\mathcal P}_d^\perp}^2 (\vect x, t), \SBtagc{25d} \label{eq:Hamiltonian_com_AL_orig} \\
	H_\text{L} &= \frac{\varepsilon_0}{2} \int\D^3\vect x \, ({\vect E^\perp}^2 + c^2 \vect B^2), \SBtag{25e} \\
	H_\text{X} &= - \frac{(\vect P \cdot \vect p_{\vect r})^2}{2 M^2 \mu c^2} + \frac{e^2}{4\pi\varepsilon_0 r} \frac{(\vect P\cdot \vect r / r)^2}{2 M^2 c^2}
		\\&\quad + \frac{m_1 - m_2}{2\mu M^2 c^2} \bigg\{ (\vect P \cdot \vect p_{\vect r}) \vect p_{\vect r}^2 / \mu - \frac{e^2}{8\pi\varepsilon_0}
		\\&\quad \times \left[\frac{1}{r} \vect P \cdot \vect p_{\vect r} + \frac{1}{r^3} (\vect P \cdot \vect r) (\vect r \cdot \vect p_{\vect r}) + \text{H.c.}\right] \bigg\}. \SBtag{25f} \label{eq:Hamiltonian_com_cross_orig}
\end{align*}
\end{subequations}

The canonical transformation into new coordinates $\vect Q,\vect q,\vect p$ used to eliminate the cross terms $H_\text{X}$ reads as follows:
\begin{subequations}
	\makeatletter
	\def\@currentlabel{\cite{sonnleitner18}.26}
	\makeatother
	\label{eq:coords_decoup_orig}
\begin{align*}
	\vect R &= \vect Q + \frac{m_1 - m_2}{2 M^2 c^2} \left[ \left(\frac{\vect p^2}{2\mu} \vect q + \text{H.c.}\right) - \frac{e^2}{4\pi \varepsilon_0 q} \vect q \right]
		\\&\quad - \frac{1}{4 M^2 c^2} [(\vect q \cdot \vect P) \vect p + (\vect P \cdot \vect p) \vect q + \text{H.c.}] \SBtag{26a} \\
	\vect r &= \vect q + \frac{m_1 - m_2}{2 \mu M^2 c^2} [(\vect q \cdot \vect P) \vect p + \text{H.c.}] - \frac{\vect q \cdot \vect P}{2 M^2 c^2} \vect P \SBtag{26b} \\
	\vect p_{\vect r} &= \vect p + \frac{\vect p \cdot \vect P}{2 M^2 c^2} \vect P - \frac{m_1 - m_2}{2 M^2 c^2}
		\\&\quad\times \left[ \frac{\vect p^2}{\mu} \vect P - \frac{e^2}{4\pi \varepsilon_0} \left(\frac{1}{q} \vect P - \frac{1}{q^3} (\vect P \cdot \vect q) \vect q\right) \right] \SBtag{26c}
\end{align*}
\end{subequations}

The internal electromagnetic potentials up to our order of approximation (thus neglecting retardation) are as follows:
\begin{align*}
	\phi_\text{el.,ng}(\vect x,t) &= \frac{1}{4\pi\varepsilon_0} \int\D^3\vect x' \, \frac{\rho(\vect x',t)}{|\vect x - \vect x'|} \SBtag{A1} \label{eq:el_pot_orig} \\
	\vect{\mathcal A}^\perp_\text{ng}(\vect x,t) &\simeq \frac{\mu_0}{4\pi} \int\D^3\vect x' \, \frac{\vect j(\vect x',t)}{|\vect x - \vect x'|} + \frac{\mu_0}{(4\pi)^2} \int\D^3\vect x'
		\\&\quad \times \int\D^3\vect x'' \, \frac{\vect x - \vect x'}{|\vect x - \vect x'|^3} \frac{\vect j(\vect x'',t) \cdot (\vect x' - \vect x'')}{|\vect x' - \vect x''|^3}\\
		&= \frac{\mu_0}{8\pi} \sum_{i=1,2} e_i \left\{ \frac{\dot{\vect r}_i}{|\vect x - \vect r_i|} + \frac{(\vect x - \vect r_i) [\dot{\vect r}_i \cdot (\vect x - \vect r_i)]}{|\vect x - \vect r_i|^3} \right\} \SBtag{A3} \label{eq:mag_pot_orig}
\end{align*}
Here we have changed the variable names of the potentials to conform to our notation; in particular we added the suffix `ng', standing for `non-gravitational'.

\newpage
%%%%%%%%%%%%%%%%%%%%%%%%%%%%%%%%%%%%%%%%%%%%%%%%%%%%%%%%%%%
\nocite{apsrev41Control} % load control commands from revtex-custom.bib (to include titles in the bibliography)
\bibliography{references,revtex-custom}

\end{document}